\documentclass[twoside]{book}
\usepackage{amsmath}
\usepackage{graphicx}
\usepackage{latexsym}
\usepackage{amsfonts}
\usepackage{amssymb}
\usepackage{bm}
\usepackage{amsbsy}
\usepackage{longtable}
\usepackage{wrapfig}
\numberwithin{equation}{chapter}
\usepackage{makeidx}
\makeindex

\begin{document}

\thispagestyle{empty}

\vspace*{\fill}

\begin{centering}

\Huge{Polymer Physics:}\\

\vspace{2ex}

\Huge{Phenomenology  of}\\

\vspace{2ex}

\Huge{Polymeric Fluid Simulations}

\vspace{2ex}
\vspace*{\fill}

\Large{George D. J. Phillies}\\

\large{Professor of Physics, Emeritus}\\

\large{Worcester Polytechnic Institute}

\end{centering}

\vspace*{\fill}



\begin{center}




\end{center}



\thispagestyle{empty}

\pagebreak

\renewcommand{\thepage}{\roman{page}}

\setcounter{page}{1}

\thispagestyle{empty}

\vspace*{\fill}

\begin{center}
\begin{large}

Dedication \\

\end{large}

\end{center}

\vspace*{\fill}

\tableofcontents



\renewcommand{\thechapter}{\roman{chapter}}
\renewcommand{\thepage}{\roman{page}}
\setcounter{page}{5}

\setcounter{chapter}{0}
\setcounter{page}{1}
\renewcommand{\thechapter}{\arabic{chapter}}
\renewcommand{\thepage}{\arabic{page}}
%

\chapter{Tests of the Rouse Model\label{rousemodel}}

George D. J. Phillies \today

[This document is a Chapter from my forthcoming volume \emph{Polymer Physics: Phenomenology of Polymeric Fluid Simulations}, which I may eventually finish.  Perhaps this chapter was unusually refractory.  One sentence summary of this Chapter:  Simulations of polymer melts conclusively demonstrate that the Rouse model is invalid in polymer melts.]

\section{Introduction}

In this Chapter, we consider simulational tests of Rouse's model\cite{ROUSErouse1953a} for polymer dynamics.  Rouse's model is widely used to describe how polymer chains move in polymeric fluids, including dilute and non-dilute solutions and polymer melts. I first note that there are two fundamental sorts of tests of the model, namely \emph{direct} tests and \emph{inferential} tests.    In direct tests, one examines what the model actual says about polymer motions, notably the Rouse mode amplitudes and their time correlation functions.  The Rouse modes either do or do not have the behavior determined by the model.  In inferential tests, one compares the model's predictions for measurable parameters with the actual behavior of the parameters.  A well-known inferential  test of the Rouse model is the dependence of the melt viscosity on the polymer molecular weight $M$.  The observed $M^{1}$ dependence for lower-molecular-weight polymers agrees with calculations based on the model, allowing one to infer that the model agrees with experiment. The power of inferential tests can be overstated. The observed $M$-dependence would only be a demonstration that polymer dynamics are Rouse-like if the prediction could be shown to be unique, i.e., the molecular weight dependence would only prove that Rouse dynamics are correct if one could show that no fundamentally different model made the same prediction. Since there actually is a fundamentally different model of polymer dynamics, the Kirkwood-Riseman model, that predicts the same $M$-dependence, claims that the $M$-dependence proves Rouse-model behavior are in fact not sustained.  The simulational studies reviewed here provide many direct tests of the Rouse model.

Our conclusions are presaged by Likhtman's observation in his review \emph{Viscoelasticity and Molecular Rheology} in \emph{Polymer Science: A Comprehensive Reference}\cite{ROUSElikhtman2012a}, where he observes ``\emph{We note that often models are studied by theoreticians just because they are analytically solvable and used by experimentalists because of availability of analytic solutions}'', leading to his conclusion ``\emph{This coupling} [GP: between Rouse modes; see below] \emph{suggests that the Rouse mode description is not very useful for entangled polymers.}'' Nonetheless, Rouse mode analysis is widely used to describe polymer motions, so tests of its validity are reviewed at length here.

The Rouse model begins by creating the Rouse coordinates, which are a discrete Fourier transform between an index $i$ that labels the position coordinates $\mathbf{r}_{i}(t)$ of the $N$ beads in a polymer chain, the chain being viewed as a line of beads, and an index $p$ that labels the $N$ Rouse coordinates $\mathbf{X}_{p}(t)$.  For more detail, see the chapter on collective coordinates.

The Rouse coordinates, in addition to being a complete orthogonal set of coordinates that specify the positions of the beads in a polymer chain, are the normal modes of Rouse's mechanical model of a polymer.  In particular, Rouse's model predicts:
\begin{enumerate}

\item the $\mathbf{X}_{p}(t)$ are normal modes of the system, so that  $\langle X_{p}(0) X_{q}(t) \rangle = 0$ if $p \neq q$ (Note that being a normal mode and being a normal coordinate are \emph{not} the same thing. One is a statement about a mathematical linear transformation; the other is a statement about the system dynamics.);

\item the ensemble-average mean-square amplitude $\langle (X_{p}(0))^{2} \rangle$ is determined by $p$, $N$, and material variables;

\item the correlation function $\langle X_{p}(0) X_{p}(t) \rangle$ of each mode decays as a pure exponential $\exp(-t/\tau_{p})$;

\item the relaxation time $\tau_{p}$ is determined by $p$, $N$, and material variables.  For larger $N/p$, $\tau_{p}$ scales as $N^{2}/p^{2}$.

\item for each mode, the bead velocities $d \mathbf{r}_{i}(t)/dt$  are always directed exactly opposite to $\mathbf{r}_{i}(t)$, i.e., they are always directed at the chain center-of-mass.

\item under shear, a polymer coil responds via an affine deformation.

\item the thermal displacements of the beads in a polymer chain are described by independent Gaussian Random Processes.

\end{enumerate}
The Rouse model also predicts mean-square chain displacements and the polymeric contribution to the solution zero-shear viscosity.  We will deal with these two predictions elsewhere.  As seen below, simulations of polymer melts systematically reject the above predictions.  As seen below, the Rouse model does not describe polymer dynamics in polymer melts.  Likhtman's suggestion that \emph{the Rouse mode description is not very useful for entangled polymers} is entirely sustained.

Many analyses begin by assuming the fundamental validity of the Rouse\cite{ROUSErouse1953a} and Zimm\cite{ROUSEzimm1956a} models of polymer dynamics, at least on some time and distance scales. It is not always recalled that the Rouse and Zimm models were preceded by the Kirkwood-Riseman model\cite{ROUSEkirkwood1948a}.  When the Kirkwood-Riseman model is mentioned at all, it tends to be treated as being much the same as the Rouse and Zimm models.  As a grain of truth, these three models all describe a single, isolated, polymer coil, not a polymer melt.

The Rouse, Zimm, and Kirkwood-Riseman models for polymer dynamics all describe a polymer chain as a line of $N$ hydrodynamically active beads, labelled $\{0, 1, 2, \ldots, N-1\}$, linked by hydrodynamically inert Hookean springs.  The beads have cartesian coordinates $(\mathbf{r}_{0}, \mathbf{r}_{1}, \ldots \mathbf{r}_{N-1})$ and hydrodynamic drag coefficients $(f_{0}, f_{1}, \ldots, f_{N-1})$.  The springs, with force constants $k_{i}$, serve to control the average distance between bonded pairs of beads. The hydrodynamic force on a bead $i$ is
\begin{equation}
   \mathbf{F}_{iH} = f_{i} (\mathbf{v}_{i} - \mathbf{\dot{r}}_{i}).
   \label{eq:ROUSEbeadhydroforce}
\end{equation}
Here $\mathbf{\dot{r}}_{i}$ is the velocity of bead $i$ and $\mathbf{v}_{i}$ is the velocity that the fluid would have had, at the location of bead $i$, if the bead were absent.

In most treatments, the bead drag coefficients $f_{i}$ are taken to have a common value $f$.  In a few studies, simulations in which some beads have extremely large drag coefficients have given physically interesting results.  When the solvent exerts a force on a bead, from Newton's Third Law that bead exerts an equal and opposite force on the solvent.  In the Zimm and Kirkwood-Riseman models, but not the Rouse model, the forces that the beads exert on the solvent create solvent flows that perturb the motion of the solvent around each of the other beads. These perturbations are the bead-bead hydrodynamic interactions, described in the Zimm and Kirkwood-Riseman models by the Oseen tensor. The polymer coils are part of a thermal system. Corresponding to the frictional forces of equation \ref{eq:ROUSEbeadhydroforce}, the fluctuation-dissipation theorem guarantees that there must on each bead be a fluctuating thermal force $\mathbf{F}_{i}(t)$, the thermal forces serving to maintain the temperature of the system.  In the Rouse model, there are no hydrodynamic interactions between the beads, so the  $\mathbf{F}_{i}(t)$ on different beads are uncorrelated.  In the presence of hydrodynamic interactions, the fluctuating thermal forces on different beads must necessarily have cross-correlations.

It is generally ignored that the Kirkwood-Riseman and Rouse-Zimm models give completely contradictory descriptions of how polymer coils move in solution and create viscous dissipation.  Consider a linear chain having $N$ beads.  In all models, the polymer coil has three center-of-mass coordinates with a center-of-mass velocity, corresponding to an average translational motion of the entire chain.

In the Rouse and Zimm models, the Rouse transformation replaces the $3N$ Cartesian coordinates with three center-of-mass coordinates and $3N-3$ internal Rouse coordinates ${X}_{p \alpha}(t)$.  Here $p \in (1, N-1)$ and $\alpha \in (x, y, z)$. In the Rouse model, each coordinate corresponds to a Rouse mode.  The $3N-3$ Rouse modes have a common feature, namely that in each mode at least some of the beads move with respect to each other.

In contrast to the Rouse and Zimm models, in the Kirkwood-Riseman model an $N$-bead polymer coil has three translational modes, describing the averaged translation of the polymer chain, and three rotational modes, describing an averaged rotation of the polymer chain.  In translational and rotational motions, the distances between the polymer beads remain constant.  There then remain $3N-6$ internal modes in which the relative positions of the beads change with time. The Rouse and Kirkwood-Riseman models thus do not agree as to how many internal modes, modes in which the beads move with respect to each other, a polymer coil has. One model says $3N-6$ modes, while the other says $3N-3$ modes.

It should have been, but was not, immediately apparent that the Rouse and Zimm models with their $3N-3$ independent internal modes are completely inconsistent with basic classical mechanics, in which an $N$-atom molecule can translate and rotate, and therefore has $3N-6$ internal degrees of freedom.  This count of the allowed number of internal coordinates is established with absolute certainty by experimental and theoretical studies of infrared and Raman spectroscopy\cite{ROUSEherzberg1945a}.  Furthermore, prominently from Raman spectroscopy of molecular crystals, the rotational modes are slow relative to most vibrational modes, so it would be incorrect to propose that the discrepancy can be hidden in a few high-frequency modes that elsewise are of no significance.

Furthermore, the models are entire opposite in their descriptions of how polymeric viscosity increments arise. Rouse assigns viscous dissipation to the polymer chain's internal modes, while denying whole-body rotation. Kirkwood and Riseman assign viscous dissipation to whole-body rotation, while neglecting internal motions of a chain as providing only secondary corrections.

\section{The Rouse Model}

Rouse\cite{ROUSErouse1953a} proposed a simple image for an isolated polymer chain, in the form of a series of beads linked by entropic springs. Each bead represents a significant segment of the polymer, so that successive beads form a Gaussianly-distributed random walk.  For a detailed discussion of the Rouse model, see the chapter on collective coordinates. The Rouse model makes it natural to use Rouse coordinates to describe polymer dynamics. It is important to emphasize the distinction between Rouse \emph{coordinates} and Rouse \emph{modes}.  The Rouse coordinates are a set of $3N$ numbers that between them describe the positions of all the beads in a polymer coil at a given time. The Rouse modes are a set of solutions of Rouse's model for polymer dynamics; they describe the motions of a polymer coil whose forces are given by the Rouse model. Rouse \emph{modes} are described naturally in terms of Rouse \emph{coordinates}, but they can equally well be written in terms of Cartesian coordinates. However, no matter what model of polymer dynamics is correct, the Rouse coordinates continue to be a valid as a set of coordinates.

The Rouse coordinates provide the normal mode solutions to the Rouse model, namely
\begin{equation}
     \langle X_{p \alpha}(t) X_{q \beta}(0)\rangle =  \delta_{\alpha \beta}  \delta_{pq} \langle (X_{p \alpha}(0))^{2} \rangle \exp(-\Gamma_{p} t),
     \label{eq:ROUSErousemodesoln}
\end{equation}
where the brackets $\langle \cdots \rangle$ indicate a thermal average, $\delta_{\alpha \beta}$ and $\delta_{pq}$ are Kronecker deltas, and where the relaxation rate $\Gamma_{p}$ satisfies
\begin{equation}
      \Gamma_{p} = \frac{12 k_{B} T}{f b^{2}} \sin^{2}\left(\frac{p \pi}{2N}\right).
\label{eq:ROUSErouserelnrate}
\end{equation}

Finally, the model predicts
\begin{equation}\label{eq:ROUSEmeansquareX}
   \langle (X_{p\alpha}(0))^{2} \rangle =  \frac{b^{2}}{8N  \sin^{2}\left(\frac{p \pi}{2N}\right)} \approx \frac{N b^{2}}{2 p^{2} \pi^{2}} .
\end{equation}
 The $X_{p\alpha}(0)$ are predicted to have independent Gaussian random distributions, so all higher moments of $X_{p\alpha}(0)$ can be calculated from $\langle (X_{p\alpha}(0))^{2} \rangle$.

In discussing equation \ref{eq:ROUSErousemodesoln}, $X_{p \alpha}(t)$ may be recognized as the instantaneous amplitude of mode $p \alpha$.  The Rouse modes are orthogonal, in the sense that within the Rouse model $\langle X_{p \alpha}(t) X_{q \beta}(0)\rangle = 0$ at all times $t$ if $\alpha \neq \beta$ or if $p \neq q$.  The statement that the modes are orthogonal at all times arises from the forces in the Rouse model, and is not equivalent to the equally-correct statement that the basis vectors of the Rouse coordinates are orthogonal. It is straightforward to make a modest modification of the Rouse model such that some modes become cross-correlated, namely one applies to the molecule a time-independent external shear field.\cite{ROUSEphillies2018a}.

In the Rouse model, the correlation functions  $\langle X_{p\alpha }(t) X_{p\alpha}(0)\rangle$ relax exponentially in time. Contrariwise, if mode relaxations are stretched exponentials in time, or have some other time dependence, then assuredly the underlying polymer dynamics are not those of the Rouse model. In each Rouse mode, most beads are displaced from their rest positions and return back to them as time advances. In the Rouse model, all beads have the same rest position; at rest, all beads are at the center of mass.   Rouse modes correspond to spatially (but not temporally) oscillatory displacements having larger or smaller wavelengths. Rouse modes are not wavelets; they do not refer to fluctuations in a single localized region of a chain. The Zimm model is substantially similar to the Rouse model, except that bead-bead hydrodynamic interactions at the level of the Oseen tensor are included in the calculation.  These hydrodynamic interactions change how rapidly each mode relaxes, but in the Zimm model the Rouse modes remain orthogonal, and continue to relax as simple exponentials.

Rouse applied his model\cite{ROUSErouse1953a} to calculate the viscosity increment created by a Rouse-model polymer.  An applied shear field, with fluid velocity along the $x$ axis, velocity gradient along the $y$-axis, and vorticity vector in the $z$ direction,  was claimed to displace polymer beads, but according to Rouse only in the $x$ direction, the direction of the velocity. Motions of the polymer beads in the $y$ and $z$ directions were claimed not to be affected by this shear field.  In the Rouse model, a polymer chain subject to an external shear field performs an affine deformation, as described more recently by deGennes\cite{ROUSEdegennes1979a}, but in the Rouse model polymer chains in a shear field do not rotate. The forces of the polymer beads on the solvent, due to their having been displaced relative to each other by the shear, as described by Rouse modes, lead in the model to viscous dissipation.

In the Rouse model every monomer bead is exposed to a Gaussian random force.  The displacements $\mathbf{r}_{i}(t) - \mathbf{r}_{j}(0)$ arise from weighted linear sums of these external forces, so $\mathbf{r}_{i}(t) - \mathbf{r}_{j}(0)$ must in the Rouse model have a Gaussian random distribution, leading to
\begin{equation}
    \langle( \exp(\imath \mathbf{q} \cdot \left(\mathbf{r}_{i}(t) - \mathbf{r}_{j}(0)\right)) \rangle  = \exp\left( - \frac{q^{2}}{6} \left\langle (\mathbf{r}_{i}(t) -  \mathbf{r}_{j}(0))^{2} \right\rangle \right)
  \label{eq:averageofRouseexponential}
\end{equation}
and therefore
\begin{equation}
    g^{(1)}(q,t) = \sum_{i=1}^{N} \sum_{j=1}^{N} \exp\left( - \frac{q^{2}}{6} \left\langle (\mathbf{r}_{i}(t) -  \mathbf{r}_{j}(0))^{2} \right\rangle \right)
  \label{eq:rouseintermediatestructurefactor}
\end{equation}
This is the Gaussian approximation for the intermediate structure factor $g^{(1)}(q,t)$; its time dependence is determined by the one- and two-particle time-dependent mean-square particle displacements.  Chong and Fuchs\cite{ROUSEchong2002a} offer a demonstration that this Gaussian approximation is theoretically appropriate. We return later to the accuracy of this approximation as tested by simulations.

There is an inferential demonstration that the Rouse model is correct at least part of the time in polymer melts.  The demonstration begins with the experimental observation that the melt viscosity of lower-molecular-weight polymers scales linearly in the polymer molecular weight $M$.  This observation is claimed to prove the presence of Rouse dynamics, namely the Rouse model predicts the observed $\eta \sim M$. The proof is an example of a logical error, namely the invalid claim that like effects prove like causes.  In order for the claim to be valid, it would be necessary to show that no other model of polymer dynamics predicts $\eta \sim M$.  However, no such proof is possible, because the Kirkwood-Riseman model with hydrodynamic interactions suppressed also predicts $\eta \sim M$, even though the Rouse and Kirkwood-Riseman models have entirely contradictory descriptions of how polymers move in a polymeric liquid when they are contributing to the liquid's viscosity.   The $M$-dependence of $\eta$ can therefore provide only negative evidence on the validity of the Rouse model, namely if $\eta \nsim M$ in some system, then the Rouse model is not valid in that system.

An interesting physical question is the possible presence of cross-correlations in Rouse mode amplitudes in physical systems, i.e., is $\langle X_{pi}(t) X_{qj}(0)\rangle$ ever non-zero for $p \neq q$ and/or $i \neq j$?  Such correlations do not exist in a Rouse-model polymer but could exist in some system that is not described by the Rouse model.  A plausible general form for a cross-time-correlation function of two Rouse modes is
\begin{equation}\label{ROUSEcrosstcfdef}
     \Phi_{pqij}(t) = \frac{\langle \mathbf{X}_{pi}(t) \mathbf{X}_{qj}(0)\rangle}{\langle (\mathbf{X}_{pi}(0))^{2} \rangle^{1/2}\langle (\mathbf{X}_{qj}(0))^{2} \rangle^{1/2}}
\end{equation}
A corresponding equation, not the most general one, for cross-correlations between the mean-square amplitudes of two Rouse modes would be
\begin{equation}\label{ROUSEcrosstcfdef4}
     \Phi^{(4)}_{pqij}(t) = \frac{\langle |\mathbf{X}_{pi}(t)|^{2} |\mathbf{X}_{qj}(0)|^{2}\rangle}{\langle (\mathbf{X}_{pi}(0))^{2} \rangle \langle (\mathbf{X}_{qj}(0))^{2} \rangle}
\end{equation}
The elaborate normalization seen here is invoked because even a variable whose typical size is small can still make a strong contribution, relative to its size, to the variable with which it is correlated. From the above definition, $\Phi_{ppii}(0) = 1$; a variable is perfectly correlated with itself.   In an equilibrium, nonchiral fluid, from reflection symmetry  $\Phi_{pqij}(0) = 0$ if $i \neq j$.  For an isolated Rouse chain \emph{in a fluid with shear}, we have previously shown from simulations that $\Phi_{ppij}(t) \neq 0$ can occur. in the presence of shear, one finds cross-correlations for $p=q$ and $i \neq j$.\cite{ROUSEphillies2018a}

In many cases, $\langle X_{pi}(t)  X_{pi}(0)\rangle$ is found simulationally to decay as a stretched exponential in time, i.e.,
\begin{equation} \label{eq:ROUSEstretchedbasic}
     \langle X_{pi}(t)  X_{pi}(0)\rangle  =  \langle (X_{pi}(0))^{2}\rangle \exp(-\left(\frac{t}{\tau}\right)^{\beta})
\end{equation}
Some authors have proposed that a characteristic time $\tau_{p}$ may be extracted from this form, by analogy with the simpler integral
\begin{equation}\label{eq:ROUSEexpaveraged}
    \tau = \int_{0}^{\infty} dt \, \exp(-t/\tau),
\end{equation}
namely
\begin{equation}\label{ROUSEtaupdef}
   \tau_{p} = \int_{0}^{\infty}  \, dt \, \exp((-\frac{t}{\tau})^{\beta})  = \frac{\tau}{\beta}  \Gamma(1/\beta)
\end{equation}
or equivalently
\begin{equation}
   \label{eq:ROUSEgammaaveraged}
   \gamma^{-1} =  \int_{0}^{\infty} dt \, \exp(- \gamma t),
\end{equation}
leading to
\begin{equation}\label{eq:ROUSEgammapdef}
\bar{\gamma} =   \left(  \int_{0}^{\infty}  \, dt \, \exp(-\alpha {t}^{\beta}) \right)^{-1},
\end{equation}
or to
\begin{equation}\label{eq:ROUSEgammapfinal}
   \bar{\gamma} = \frac{\alpha^{1/\beta}}{ \Gamma(1+1/\beta)},
\end{equation}
where here $\Gamma(1/\beta)$ is the gamma function.  There does not appear to be a stated physical basis for identifying $\tau_{p}$ rather than some other average over $\langle X_{pi}(t)  X_{pi}(0)\rangle$ as the appropriate characteristic time, but this average is simple.

The effective relaxation rate is for the same stretched exponential in time is\cite{ROUSEpadding2002a}
\begin{equation}\label{eq:ROUSEweffective}
    W^{\rm eff}  = \frac{1}{4 \tau_{p} \sin^{2}(p \pi/2N)} \sim  \frac{T}{f b^{2}}.
\end{equation}
For the Rouse model, $W^{\rm eff}$ only depends on the temperature, the monomer friction factor $f$, and the segment length $b$.

\section{Simulation Tests of the Rouse Model\label{ROUSEextensions}}

In all solution models, viscous dissipation occurs because the polymer's beads cannot move, at every point, with exactly the velocity that the solvent would have had at the same location, if the beads were not present, leading to frictional drag and dissipation. Beyond this point, the models do not agree as to the effect of an applied shear on a polymer coil.   In the Rouse-Zimm models, the $3N-3$ internal modes naturally partition into three sets of $N-1$ modes. Each set of $N-1$ modes refers to bead displacements parallel to one of the three Cartesian coordinate axes. If a simple shear is applied, with $d v_{x}/dy$ being the non-zero constant shear, the shear velocity in the $x$ direction was assumed by Rouse to induce bead velocities parallel to the $x$ axis. Rouse assumed that no motion was induced parallel to the $y$ or $z$ axes.  The shear field thus acts on the Rouse modes parallel to the $x$ direction, while leaving the modes involving displacements parallel to the $y$ and $z$ directions unperturbed. Rouse's formula for the polymeric viscosity increment is based on this assumption.  Furthermore, under the influence of a shear field, the Rouse relaxation modes were claimed to continue to satisfy equation \ref{eq:ROUSErousemodesoln}, so that the $X_{p}$ continue to be uncorrelated, while the relaxation rates remained independent of the applied shear rate.

In contrast, in the Kirkwood-Riseman model, under the influence of a shear field a polymer coil is assumed to rotate. An applied shear field $d v_{x}/dy$ creates bead motions parallel and antiparallel to the $x$ axis. However, because the polymer response is rotational, this shear field creates an equal amount of bead motion parallel and antiparallel to the $y$ axis.  Kirkwood and Risemann approximated the polymer internal modes, or at least their response to shear, to be negligible.  The model of Kirkwood and Riseman does not consider the response of polymer internal modes to an applied shear.

To resolve this contradiction between the Rouse and Kirkwood-Riseman models, Phillies\cite{ROUSEphillies2018a} made Brownian dynamics simulations on a single bead-spring polymer coil. Hydrodynamic interactions were not included, so the random thermal forces on separate beads could be treated as being uncorrelated. As the test was of the Rouse model, the original Rouse potential
\begin{equation}
       U_{r} =  \sum_{i=1}^{N-1} \frac{1}{2} k (r_{i,i+1})^2
\label{eq:ROUSErousepot}
\end{equation}
 was applied to the beads. Here the sum is over all pairs of bonded beads.  $k$ is a spring constant, while $r_{a,b}$ is the scalar distance between covalently linked beads $a$ and $b$. Chains had no bending constraints.  Remote parts of the chain were able to pass ghostlike through each other.

Phillies established that the Kirkwood-Riseman model, so far as it goes, is correct, while the Rouse and Zimm  models are wrong for a polymer coil in a shear field.  In particular, these simulations demonstrated for coils in a shear field:  Polymer coils do indeed rotate, so that
\begin{equation}
      \left\langle \sum_{i=1}^{N}  y_{i}  v_{xi} \right\rangle = - \left\langle \sum_{i=1}^{N}  x_{i}  v_{yi} \right\rangle.
\label{eq:theyrotate}
\end{equation}
Here $x_{i}$ and $y_{i}$ are the $x$ and $y$ components of bead $i$'s location relative to the polymer center of mass, while $v_{xi}$ and $v_{y i}$ are the $x$ and $y$ components of bead $i$'s velocity. In shear, Rouse amplitudes become cross-correlated, so the Rouse coordinates cease to represent normal modes. The mean-square amplitudes and relaxation rates of Rouse modes depend on the shear rate.  The error in the Rouse and Zimm models is at their very beginning.  Their equations of motion for the polymer beads have no applied shear, so the model only refers to an isolated polymer coil in a quiescent liquid.  Rouse-Zimm coils therefore do not rotate.  However, they are also not subject to a shear field, so they do not create viscous dissipation.

Arising from these theoretical models is the question of what one means by a 'bead'.  Polymer molecules are not actually formed from little spheres connected by very thin Hooke's-law springs. The beads and springs are abstracted from an actual description of a polymer molecule.  Are the beads entirely an abstraction, or do they have some meaningful size?

Agapov and Sokolov\cite{ROUSEagapov2010a} compare various implicit determinations of bead size with the nominal Kuhn length $b$. In the Kuhn model, a polymer is divided into $N$ segments of length $b$, the segments being straight and `freely-jointed', i.e., each segment is free to make an arbitrary angle with the next segment in line. $N$ and $b$ are determined by two constraints, namely that the length of the fully-stretched polymer is $Nb$, while the root-mean-square polymer end-to-end distance $R$ satisfies $\langle R^{2} \rangle = Nb^{2}$.  Agapov and Sokolov note that these definitions of $N$ and $b$ refer purely to polymer statics, but that a bead of the dynamic Rouse model has often been identified with a Kuhn segment.  The notion of the identification was that the Kuhn length was the length of the shortest chain segment that followed Rouse dynamics, at least in the melt, and that the dynamics of shorter chain segments were not described by the Rouse model. As Agapov and Sokolov explain, for a Rouse chain the relaxation rate of the dynamic structure factor $S(q,t)$, typically as obtained from neutron scattering, scales as $q^{4}$.  However, if the scattering vector $q$ is made sufficiently large, one is probing chain motions over distances less than that for which the Rouse model is valid, in which case at some $q = Q$ the relaxation rate deviates from $q^{4}$ behavior.  The corresponding distance $2 \pi/Q$ defines a dynamic bead size.  Agapov and Sokolov note the analysis of Nicholson, et al.,\cite{ROUSEnicholson1981a} that in PDMS $2 \pi/Q \approx b \approx 16\text{\AA }$ but that in polystyrene  $2 \pi/Q \approx 55\text{\AA }$ is about 2.5 times as large as $b$, i.e., the dynamic bead size inferred from $S(q,t)$ is perhaps 2.5 times the Kuhn length.  Agapov and Sokolov also note computer simulations\cite{ROUSEpaul1998a} and oscillatory flow birefringence studies\cite{ROUSEamelar1991a} that found a dynamic bead size that is considerably larger than the Kuhn length.

Colmenero\cite{ROUSEcolmenero2015a} observes that the two fundamental results of the Rouse model are that the Rouse amplitudes are independent, so that $\Phi_{pqij}(0) = 0$ if $p \neq q$, and that the Rouse correlators decay exponentially, namely $\Phi_{ppii}(t) \sim \exp(-t/\tau)$.  He notes a variety of cases in which the observed correlators relax as stretched exponentials rather than exponentials in time, for example in cold melts\cite{ROUSEarreseigor2014a} and in blends\cite{ROUSEmoreno2008a,ROUSEbrodeck2010a,ROUSEarreseigor2011a}. As an interpretation, he suggests that, at low temperatures, non-exponential decay of Rouse correlators might arise from coupling to local density fluctuations (the $\alpha$ relaxation).   Issues then arise from the non-exponential time dependence of the Rouse amplitude correlation functions $\Phi_{ppii}(t)$.  He proposes that various time-dependent physical quantities, such as coherent and incoherent scattering functions and dielectric relaxation spectra, have therefore been calculated incorrectly because the calculation invoked an assumed, but non-existent, exponential relaxation of the $\Phi_{ppii}(t)$.

Colmenero\cite{ROUSEcolmenero2015a} proposed to interpret the non-exponential dependence of $\Phi_{ppii}(t)$ by replacing the Rouse model's Langevin equation of motion with a Generalized Langevin Equation. Here $p$ and $q$ are mode numbers, while $i$ and $j$ label the three Cartesian coordinates.  All relaxations are pure exponentials.  Cross-correlation functions $\langle X_{pi}(t) Y_{qj}(0)\rangle$ with $p \neq q$ and/or $i \neq j$ all vanish. The relaxation times $\tau_{p}$ are
\begin{equation}
    \frac{d \Phi_{ppii}(t)}{dt} + \frac{1}{f}\int_{0}^{t} \, ds \, \Gamma(t-s)  \frac{d \Phi_{ppii}(s)}{ds} = - \frac{1}{\tau_{p}} \Phi_{ppii}(t)
      \label{eq:ROUSEcolmeneroGLE}
\end{equation}
in which $\Gamma(t-s)$ is a memory function.  (An integration by parts would replace the time derivative of $\Phi_{ppii}(s)$ with the function itself.)  To solve this equation, $\Gamma(t-s)$ was assumed to be short-lived, so that the convolution integral became nearly a single-time product, so on defining $\xi(t) = \int_{0}^{t} ds \Gamma(s)$, an approximate solution was proposed to be
\begin{equation}
     \Phi_{ppii}(t) =  \Phi_{ppii}(0)  \exp \left(-\frac{f}{\tau_{p}}\int_{0}^{t}  ds \frac{1}{f+\xi(s)} \right)
\end{equation}
On requiring this equation to yield a stretched exponential in $t$, Colmenero proposes that, in the time regime in which $\xi(t) \gg f$,  the simplest form for $\xi(s)$ is a power law in $s$. He uses his results to derive a stretched-exponential form for the time autocorrelation function $\langle \mathbf{R}_{E}(0) \cdot \mathbf{R}_{E}(t) \rangle$ of the polymer end-to-end vector $\mathbf{R}_{E}$. Comparison was then made with prior atomistic molecular dynamics simulations\cite{ROUSEbrodeck2010a,ROUSEbrodeck2009a} of polyethylene oxide and PMMA/PEO melts.  $\langle \mathbf{R}_{E}(0) \cdot \mathbf{R}_{E}(t) \rangle$  and the self part of the dynamic structure factor at a series of temperatures were then successfully fit to the predicted stretched-exponential time dependences. Values of $\tau_{p}$ and $\beta$ at each temperature, as obtained from the two physical quantities, were said to be in rather good agreement. Colmenero also compared with the Ngai coupling model.\cite{ROUSEngai1979a,ROUSEngai1996a}.

A rarely-tested prediction of the Rouse model states that Rouse modes are orthogonal in the sense that $\langle X_{p}(t) X_{q}(0)\rangle = 0$ if $p \neq q$. There have been several tests of this prediction for the special case $t=0$, including results of Kopf, et al.,\cite{ROUSEkopf1997a} and Tsalikis, et al.\cite{ROUSEtsalikis2017a}.  However, as part of an extended review of theoretical models for viscoelasticity and molecular rheology, Likhtman\cite{ROUSElikhtman2012a} obtained $\langle X_{1}(t) X_{3}(0) \rangle$ for several models of a melt of entangled polymers.  While $\langle X_{1}(0) X_{3}(0) \rangle \approx 0$ to good approximation, with increasing $t$ Likhtman found (his Figure 33) that $\langle X_{1}(t) X_{3}(0) \rangle$ increases substantially with increasing $t$, to far above any noise in the simulation, and then fades away.  Likhtman's result is entirely contrary to expectations from Rouse model dynamics, in which modes are not cross-correlated at any time, leading Likhtman to his observation as quoted above that   "\emph{This coupling suggests that the Rouse mode description is not very useful for entangled polymers.}" In citing this result, Kalathi, et al.,\cite{ROUSEkalathi2014a} nonetheless used a Rouse mode analysis in their work, saying as a sensible defense of their analysis that experimentalists ``...still tend to model chain dynamics in the language of the Rouse model.  Understanding experimental results therefore requires us to analyze the simulations in the same manner."
    `
\section{Simulations of Rouse Modes\label{ROUSElinear}}

Dynamic simulations of polymers are readily traced back to the work of Grest and Kremer\cite{ROUSEgrest1986a}, who simulated a bead-spring model for a polymer chain, in which the beads are subject to independently fluctuating thermal forces, all bead pairs separated by less than a specified distance interact with a Lennard-Jones potential, and each bead's motions are coupled to a heat bath that supplied a friction term and a thermal driving force. The interaction between bonded beads was represented with a finitely extensible (FENE) potential
\begin{equation}\label{eq:ROUSEfenepot}
    U_{ij}(r_{ij}) - 0.5 k R_{o}^{2} \ln(1 - (r_{ij}/R_{o})^{2}),
\end{equation}
where $r_{ij}$ is the distance between two beads, $k$ and $R_{o}$ are simulational parameters, and the potential is set to zero for $r_{ij} > R_{o}$.  The above potential is not the harmonic potential used by Rouse, so strictly speaking this simulation was not a test of the Rouse model. Single chains and rings containing 50-200 beads, and a 200-bead chain with no Lennard-Jones potentials, were examined. The diffusion coefficient, bead mean-square displacement, mean-square center-of-mass displacement, and static structure factor $S(q)$ were calculated.   Comparisons were made with theoretical expectations.

Kremer and Grest\cite{ROUSEkremer1990b} then reported their pioneering study of a melt of bead-spring polymers.  They made a molecular dynamics simulation in which all beads exerted a purely repulsive Lennard-Jones potential and had an attractive FENE potential between next neighbors along each polymer chain, a weak frictional force $-f\mathbf{v}$, and a corresponding thermal force. Systems with chain lengths $N$ from 5 to 400 beads and a total of 250 to 20,000 beads, corresponding to 16 to 100 chains, were studied. The nominal entanglement length was reported to be $\approx 35$ beads, so these simulations included unentangled and entangled systems.  Static properties including the mean-square end-to-end distance, the radius of gyration, the static structure factor, and the mean-square amplitude of Rouse modes were examined; these quantities showed the expected scaling dependences on $N$.  Time-dependences of the mean-square displacements of single monomers, chain centers-of-mass, and monomers relative to the center of mass, of Rouse mode amplitudes, of scattering functions, and of chain motion relative to a primitive path were also analyzed.  This chapter focuses on the Rouse modes.

\begin{figure}[thb]
\centering\includegraphics[scale=0.3]{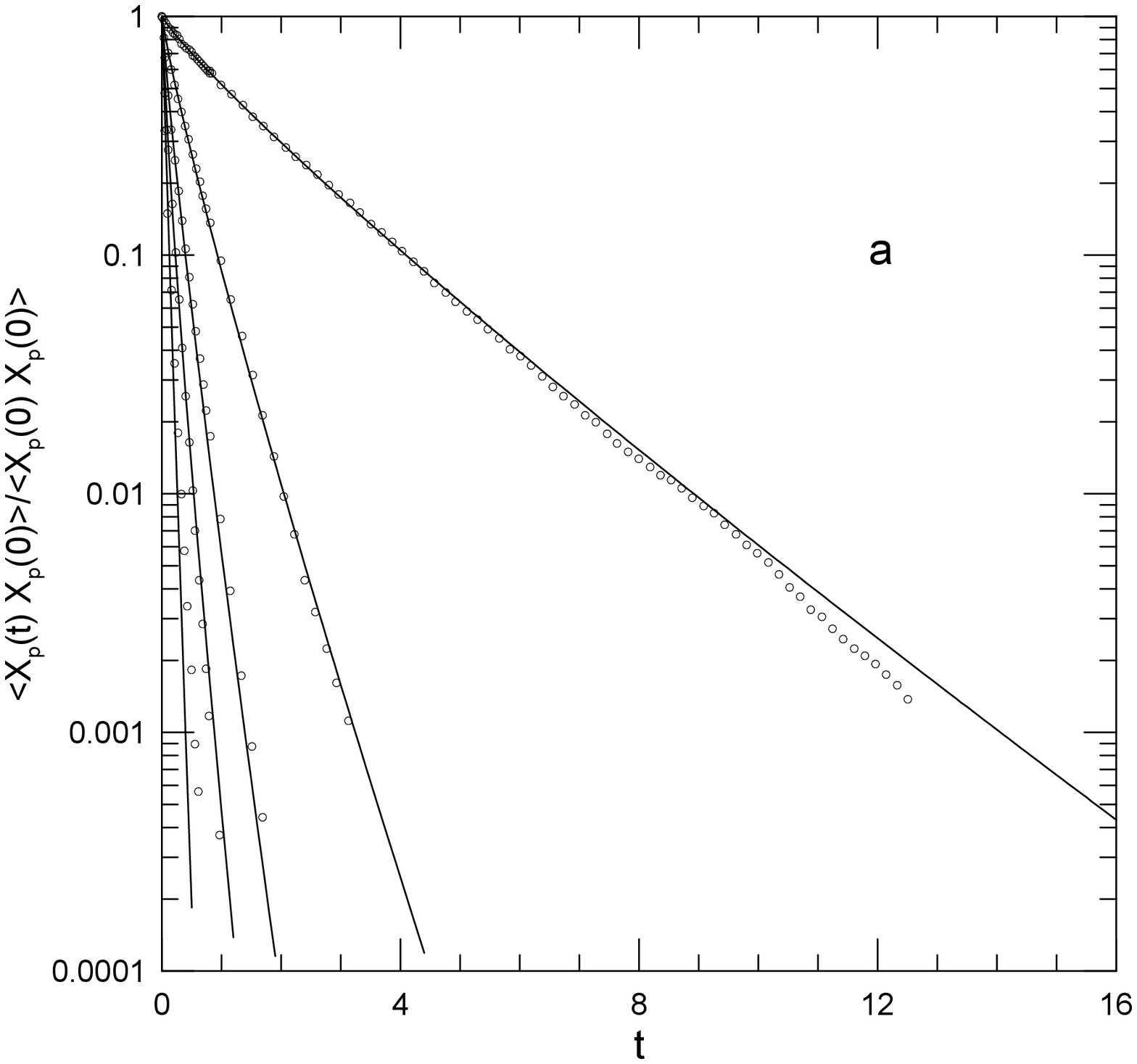}
\centering\includegraphics[scale=0.3]{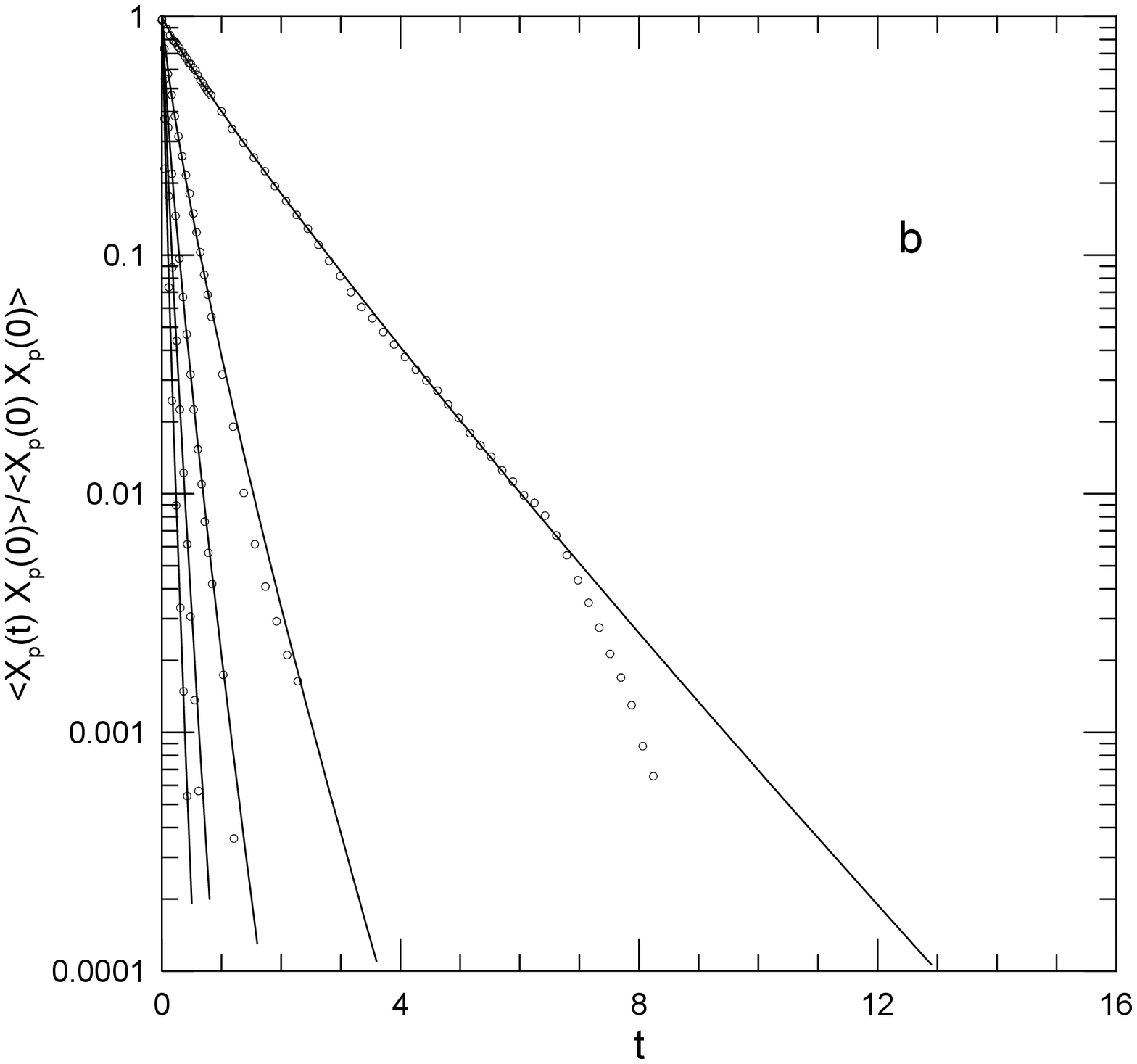}
\centering\includegraphics[scale=0.3]{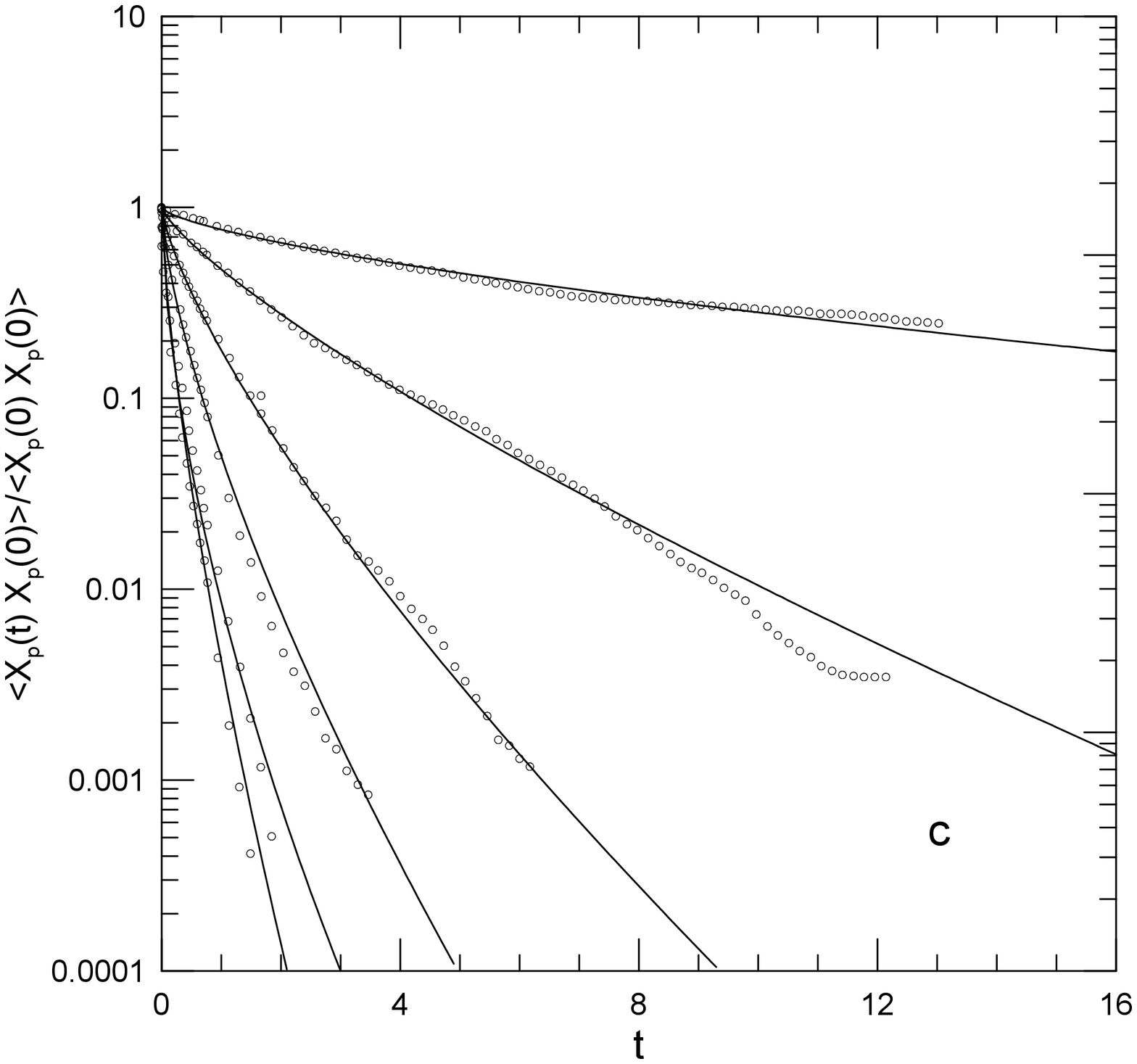}
\centering\includegraphics[scale=0.3]{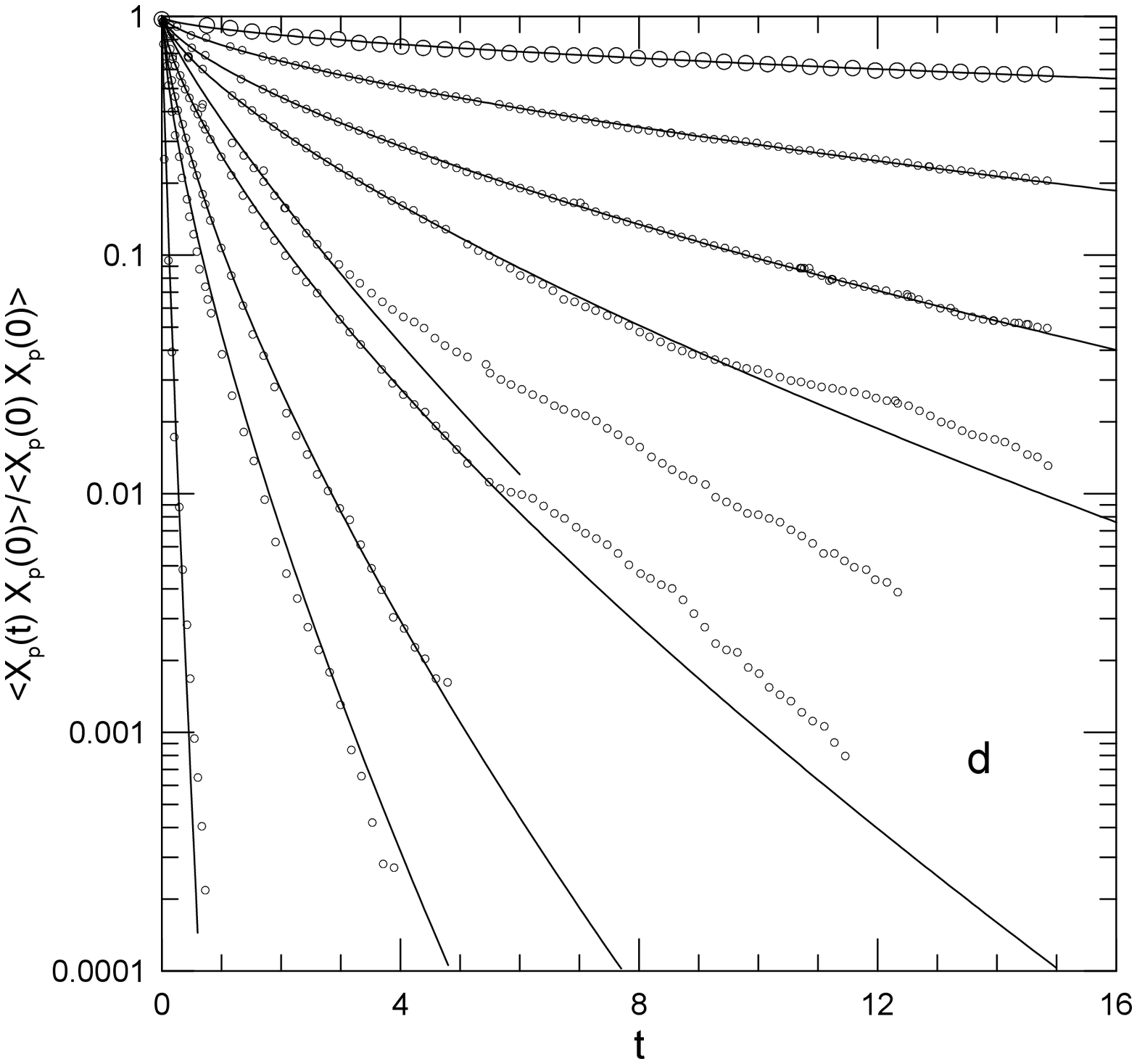}
\caption{Time autocorrelation functions for Rouse mode amplitudes $X_{p}(t)$ for $p = 1, 2, 3, 4, 5,6, 8, 10, 20$ (top to bottom) for melts of bead-spring linear polymers having $N$ of (a) 20, (b) 50, (c) 100, and (d) 200. Larger values of $p$ do not appear in all graphs; values of $p$ are listed in Table \ref{ROUSEkremer1990bTCFtable}. Circles represent the original results of Kremer and Grest\cite{ROUSEkremer1990b}; solid lines are stretched exponentials.\label{ROUSEfigkremer1990b13}}
\end{figure}

Kremer and Grest reported the normalized Rouse mode temporal autocorrelation functions
\begin{equation} \label{eq:ROUSErousentcfs}
      \frac{\langle X_{p}(t)X_{p}(0) \rangle}{\langle(X_{p}(0))^{2} \rangle}
\end{equation}
for chain lengths $N = 20, 50, 100,$ and $200$.  Figure \ref{ROUSEfigkremer1990b13} shows their results together with fits to stretched exponentials in time.  In these pioneering studies, for the two longer chains, at long times the correlation functions show weak fluctuations on top of the stretched exponentials, making it difficult to determine $\nu$ accurately.  Kremer and Grest also evaluated $\langle X_{p}(0)X_{q}(0) \rangle$ for $p \neq q$, finding that the static cross-correlations vanish within the noise in the simulations.

\begin{table}
  \centering
\begin{tabular}{|r|l|l|l|l|l|l|l|l|}
\hline
$N$ & 20 & 20  &50&50&100&100&200&200 \\ \hline
$p$ & $\alpha$ & $\beta$  & $\alpha$ & $\beta$ & $\alpha$ & $\beta$ & $\alpha$ & $\beta$ \\ \hline
1  &   0.66  &   0.89 &    0.91  &   0.90 &    0.27  &   0.68 &  0.120  &   0.58\\
2  &   2.45  &   0.88*&    3.10  &   0.75  &    0.27  &   0.68 &  0.27  &   0.65 \\
3  &    5.18  &   0.87 &    6.16  &   0.79  &    1.72   &   0.75*&   0.49  &   0.68\\
4  &    7.66  &   0.81 &    10.26  &   0.84  &    3.00  &   0.7* &   0.68  &   0.71\\
5  &    16.63  &   0.95&    15.16  &   0.83&    3.00  &   0.7*   &   1.00   &   0.83**\\
6  & & & & &    3.00  &   0.7*&   1.34  &   0.71* \\
8  & & & & & & &   2.23  &   0.69 \\
10 & & & & & & &  3.05  &   0.75* \\
20 & & & & & & &  13.65 & 0.85* \\  \hline
\end{tabular}
  \caption{Fitting parameters $\alpha$ and $\beta$ from $\langle X_{pi}(t) X_{pi}(0) \rangle/\langle (X_{pi}(0))^{2} \rangle = \exp(- \alpha t^{\beta})$, using results of Kremer and Grest\cite{ROUSEkremer1990b}, Figure 13, of N = 20, 50, 100, and 200 bead chains in their melts. In some cases, marked *, $\beta$ was fixed.  **fit was only to the points with $t < 5$ }\label{ROUSEkremer1990bTCFtable}
\end{table}

Tsalikis, et al.\cite{ROUSEtsalikis2017a} report simulations of ring polymers.  Their study is noteworthy for the range of chain parameters that were studied during the course of their simulations. A major focus of the work is comparison with Rouse model predictions for chain dynamics, but a considerable number of other parameters were also studied.  These workers report an extended series of molecular dynamics simulations of 5, 10, and 20 kDa poly(ethylene oxide) ring polymers in the melt, corresponding to polymers having 120, 227, or 455 monomers. Simulations were made with a united-atom force field\cite{ROUSEfischer2008a,ROUSEfischer2008b} under isothermal/isobaric conditions, with $T=413$ K and $P = 1$ atm. The force field parameters were expected to be sufficiently realistic that quantitative comparisons with experiments were expected to be possible, as confirmed in the paper. For the largest polymer, the simulation cell contained more than 50,000 atoms, the simulation being extended to an equivalent of 2.2 $\mu$S.

In considering Tsalikis, et al.'s findings on the applicability of the Rouse model to ring melts, one might say that the cup is half full or half empty.  Tsalikis, et al., chose to emphasize points where their simulations clearly match Rouse's predictions.  Here we emphasize the differences, points where the simulations do not match the Rouse model as applied to a ring polymer.

Tsalikis, et al., use their simulation data to compute for their rings the mean-square Rouse amplitudes.  The Rouse model predicts that the normalized amplitude $\langle (X_{p\alpha}(0))^{2} p^{2}/N$ should be independent of mode number $p$ and polymer bead count $N$.  This prediction is rejected by Tsalikis, et al.'s, simulations:  The normalized amplitudes depend on $p$, and  at small $p$ are smaller than predicted by the Rouse model. For the $N=455$ polymer, the normalized amplitude for $p=2$ is modestly more than half its value for the same polymer at large $p$.  The range of smaller $p$-values for which the normalized amplitudes are below their large-$p$ limit increases with increasing $N$, the increase in the range being approximately linear in $N$.  However, for all $N$ studied, the normalized amplitudes do appear to go to the same large-$p$ limit, so the model is arguable valid for large $p$.

\begin{figure}[thb]
\centering\includegraphics[scale=0.3]{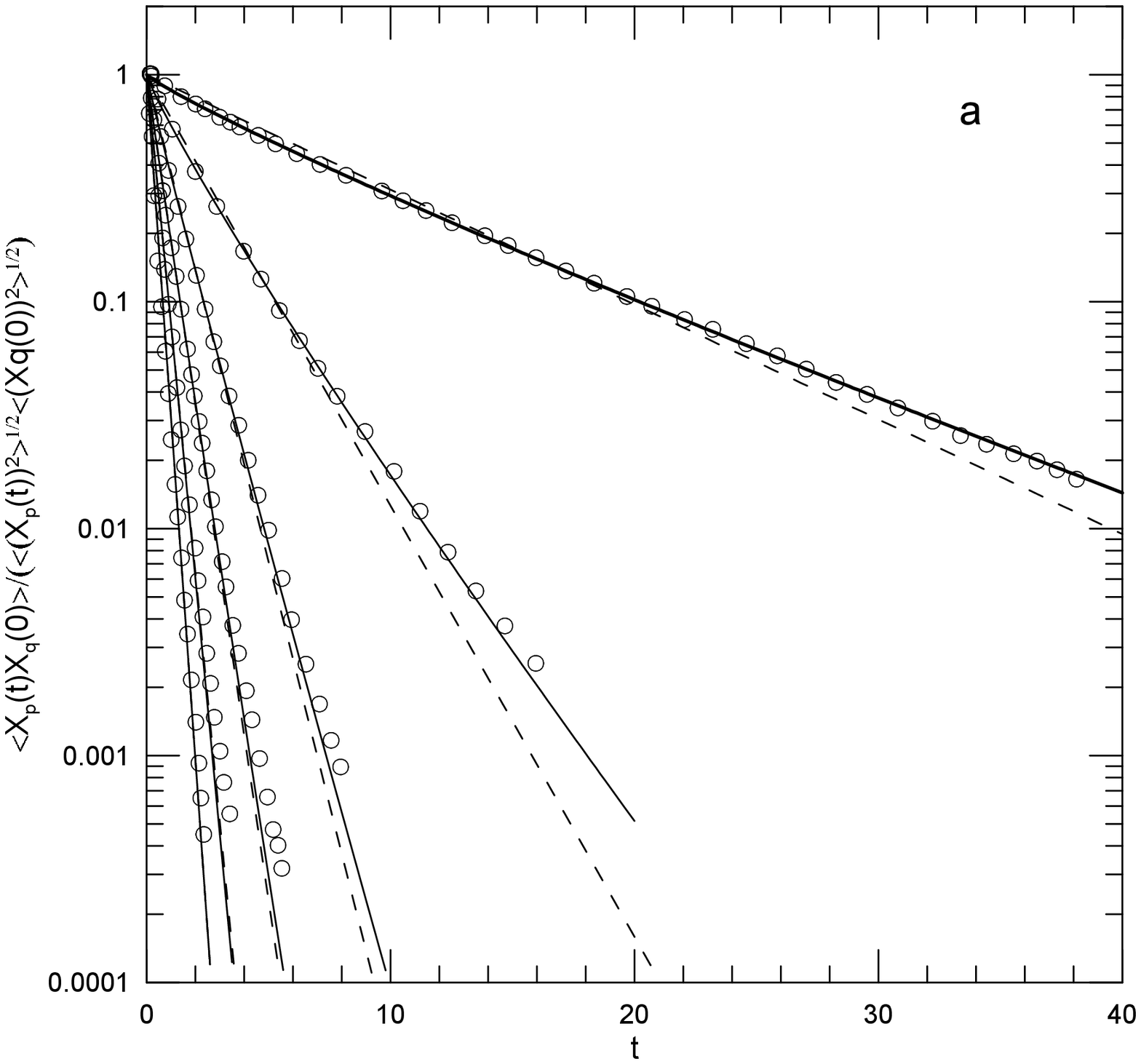}
\centering\includegraphics[scale=0.3]{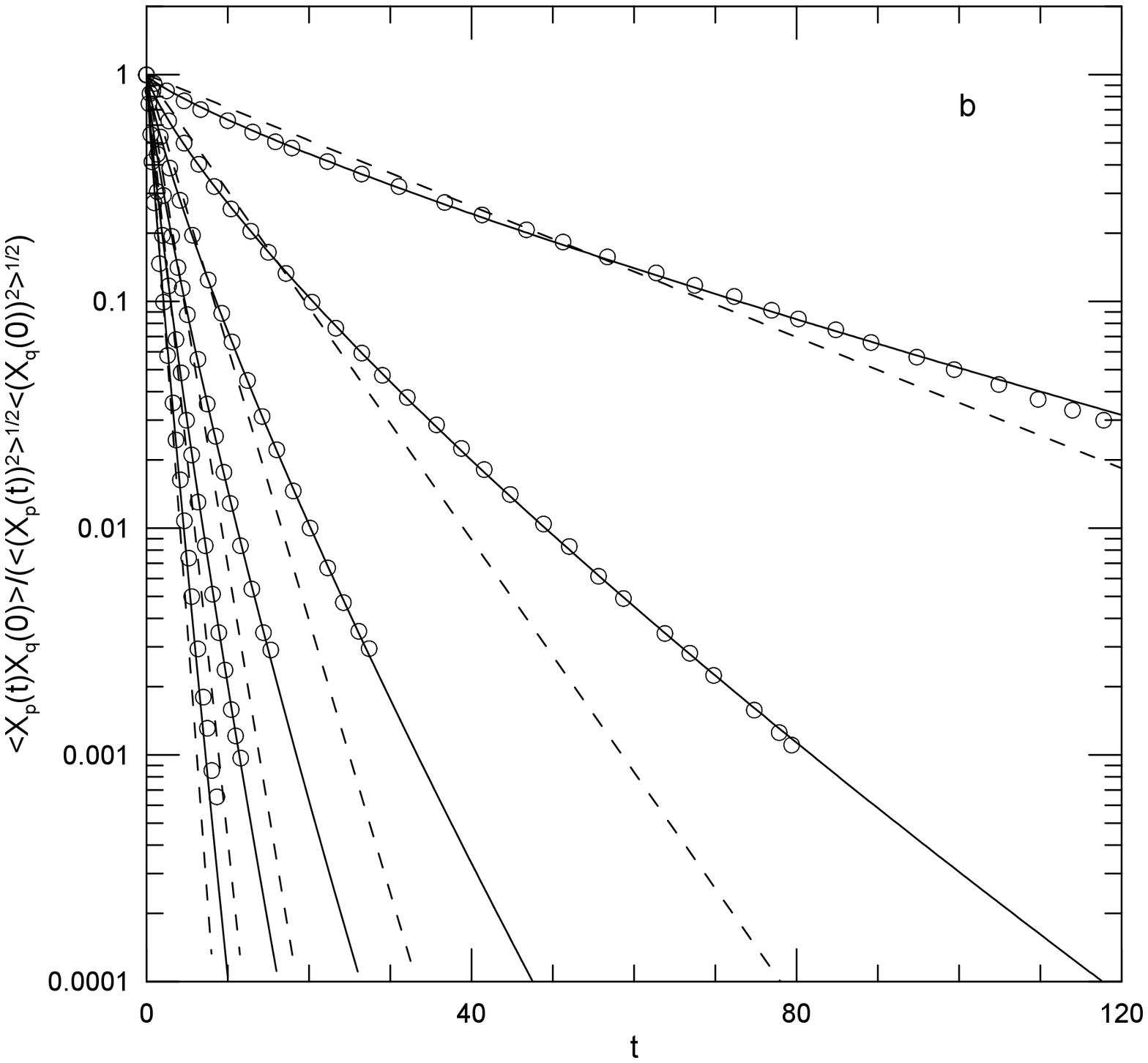}
\centering\includegraphics[scale=0.3]{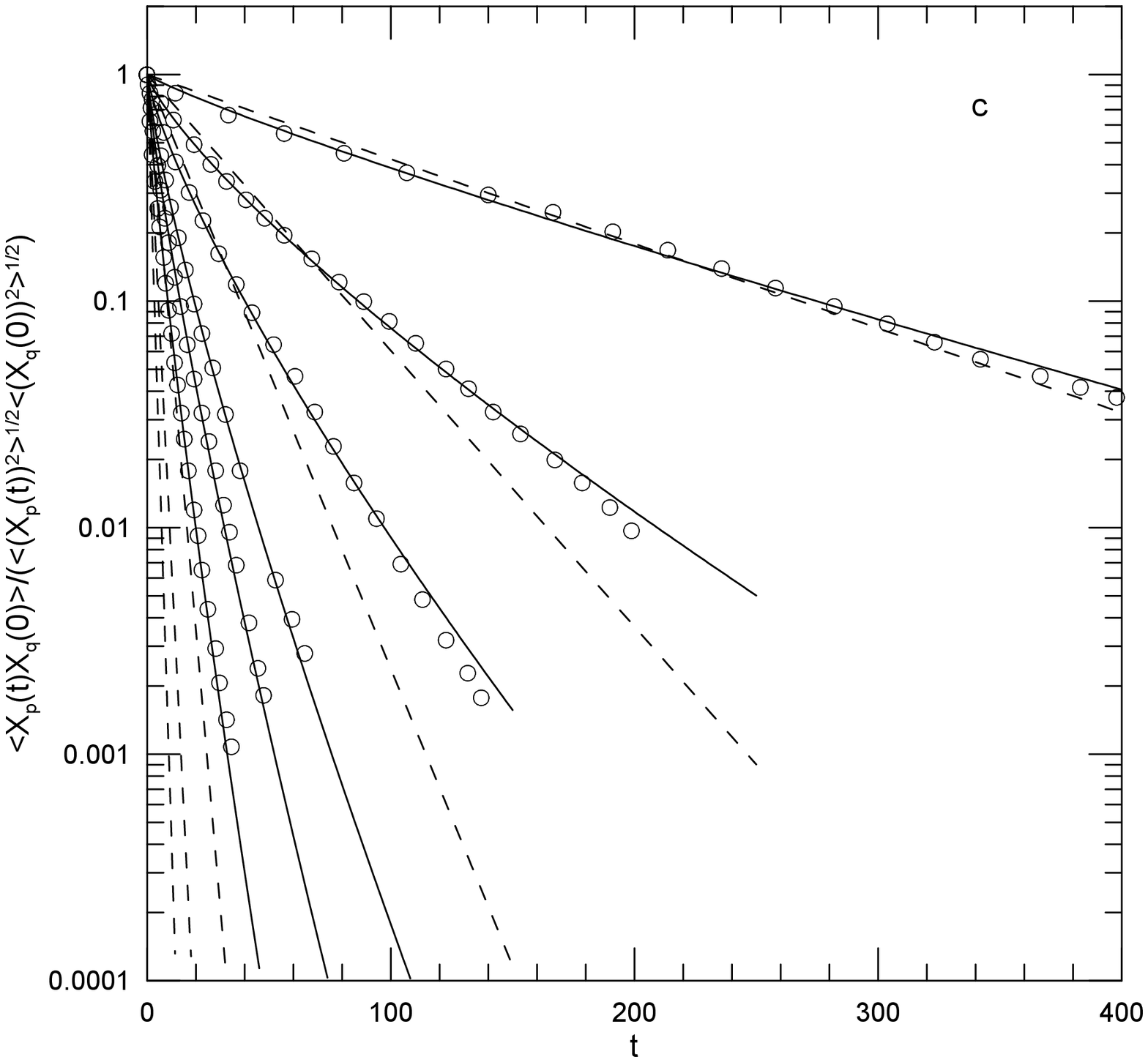}
\centering\includegraphics[scale=0.3]{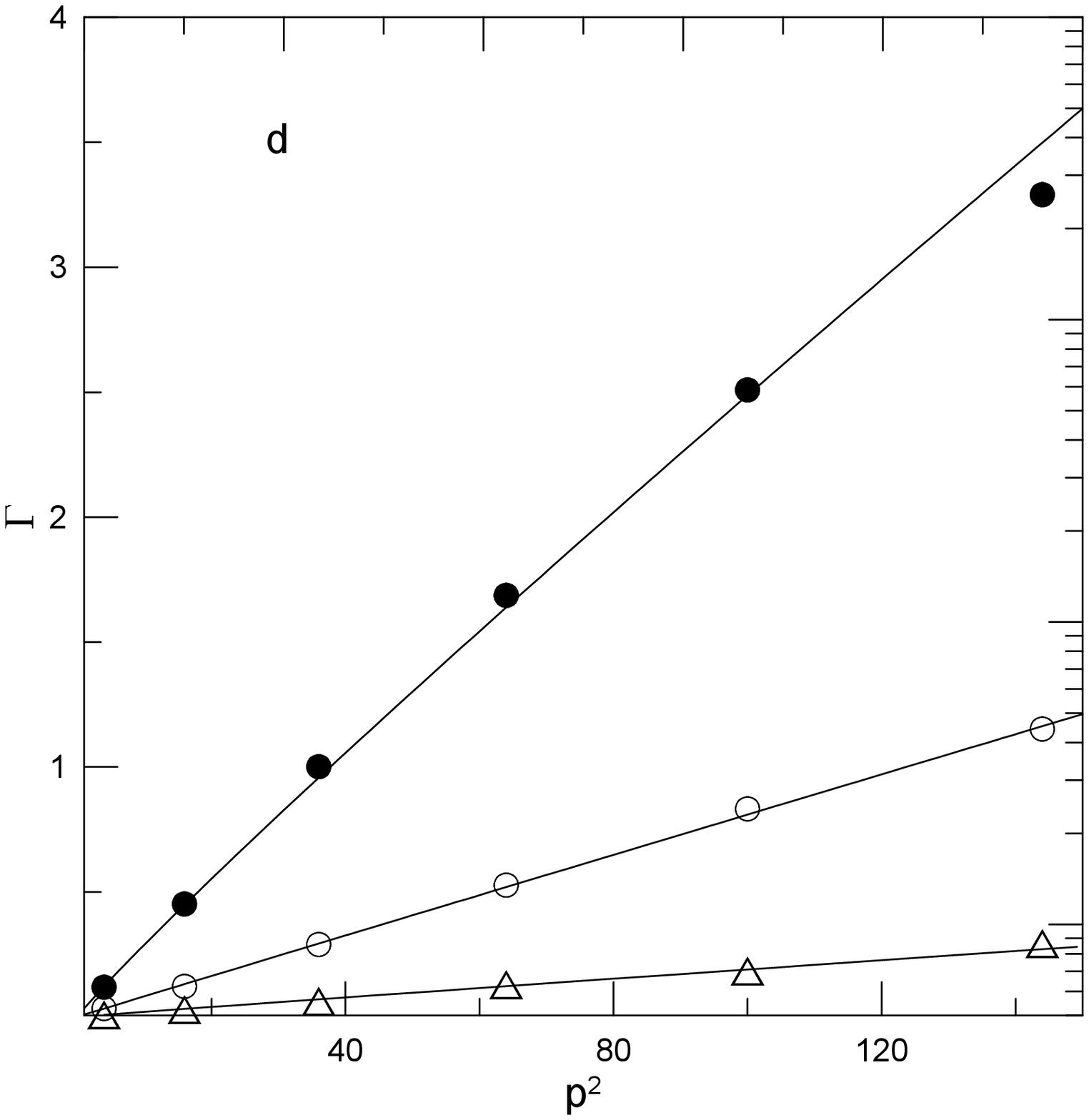}
\caption{Time autocorrelation functions for Rouse mode amplitudes $X_{p}(t)$ for $p =  2, 4, 6, 8, 10, 12$ (top to bottom) for melts of  (a) 5 kDa, (b) 10 kDa, and (c) 20kDa poly(ethylene oxide) ring polymers. (d) shows $\Gamma$ against $p^{2}$.  Lines are power-law fits to $\Gamma = A p^{2x}$ for $x = 0.94, 0.99$, and $0.98$ (top to bottom). Circles represent samplings of points from the original results of Tsalikis, et al.\cite{ROUSEtsalikis2017a}; dashed lines are  best-fits to a single exponential; and solid lines are non-linear least-squares fits to stretched exponential. fits.\label{ROUSEfigtsalikis2017aSI8}}
\end{figure}

For each of their chain lengths and $p = 2, 4, 6, 8, 10,$ and $12$, Tsalikis, et al., also report the time dependence of the time correlation functions $C(t) = \langle X_{pi}(t) X_{pi}(0) \rangle$ .  Figure \ref{ROUSEfigtsalikis2017aSI8} shows a sampling of their measurements(dots).  The figure also shows our fits of these dots to stretched exponentials (solid lines)
\begin{equation}\label{ROUSEstretchedt}
   \langle X_{pi}(t) X_{pi}(0) \rangle/\langle (X_{pi}(0))^{2}\rangle  = \exp( - \alpha t^{\beta})
\end{equation}
and to pure exponentials (dashed lines, fits to the initial slope).  Here $\alpha$ and $\beta$ are fitting parameters. The correlation functions were normalized to unity at $t=0$.  If Figure \ref{ROUSEfigtsalikis2017aSI8} is examined, it is apparent that the relaxation of $\langle X_{pi}(t) X_{pi}(0) \rangle$ is described well by a stretched exponential in time, except for a few of the largest-$t$ points, contrary to the Rouse model prediction that the relaxations should be simple exponentials.

The stretched exponential is characterized by $\alpha$ and $\beta$. As an alternative, we follow equation \ref{ROUSEtaupdef} and define an average decay rate $\bar{\gamma}$ as $\bar{\gamma} \equiv \tau_{p}^{-1}$  where here $\Gamma(s)$ is the Gamma function.  Numerical values for the fitting parameters and the computed $\bar{\gamma}$ are seen in Table \ref{ROUSEtsalikis2017aTCFtable}.  For each molecular weight, $\alpha$ increases nearly 40-fold between $p$=2 and $p=12$.  $\beta$ is close to unity for the $N=120$ polymer, but about 0.8 for the two larger rings.

\begin{table}
  \centering
\begin{tabular}{|r|r|l|l|l|l|}
$p$ &$M$(kDa) & $\gamma_{1}$ $\alpha$ & $\beta$ & $\bar{\gamma}$ \\ \hline
2 &	5&	0.117&	0.157&	0.894&	0.119\\
4&	5&	0.437&	0.513&	0.898&	0.452\\
6&	5&	0.988&	1.019&	0.958&	1.001\\
8&	5&	1.675&	1.683&	0.978&	1.687\\
10&	5&	2.524&	2.527&	1.014&	2.510\\
12&	5&	3.471&	4.230&	1.297&	3.291\\ \hline
2&	10&	0.033&	0.070&	0.814&	0.0341\\
4&	10&	0.118& 0.210&	0.793&	0.123\\
6&	10&	0.277& 0.403&	0.810&	0.290\\
8&	10&	0.498& 0.658&	0.807&	0.528\\
10&	10&	0.777& 0.942&	0.818&	0.833\\
12&	10&	1.116& 1.196&	0.884&	1.152\\ \hline
2&	20&	0.00859& 0.0166&	0.878&	0.00882\\
4&	20&	0.028& 0.0690&	0.786&	0.0290\\
6&	20&	0.060& 0.128&	0.782&	0.0629\\
8&	20&	0.126& 0.213&	0.804&	0.129\\
10&	20&	0.181& 0.276&	0.813&	0.184\\
12&	20&	0.282& 0.404&	0.813& 	0.293 \\  \hline
\end{tabular}
  \caption{Fitting parameters $\alpha$ and $\beta$ and average decay rates $\bar{\gamma}$ from $\langle X_{pi}(t) X_{pi}(0) \rangle$ based on simulations of Tsalikis, et al.,\cite{ROUSEtsalikis2017a} of 5, 10, and 20 kDa polyethylene oxide ring melts. $\gamma_{1}$ is the initial slope, corresponding to a simple exponential relaxation. }\label{ROUSEtsalikis2017aTCFtable}
\end{table}

Tsalikis, et al., make the valuable and correct point that $C(t)$ as displayed on a semilog plot 'seems to be exponential-like' (i.e., is close to a straight line), except at short times.  This point does not contradict our observation that $C(t)$ follows well a stretched exponential in time. 'Exponential-like' behavior is a general feature of stretched-exponential time dependences on semilog plots, namely if we have a function
\begin{equation}\label{ROUSEderivative}
   y = \exp(- \alpha t^{\beta})
\end{equation}
then its logarithmic derivative is
\begin{equation}\label{ROUSEderivative2}
  \frac{d \ln(y)}{dt} = - \alpha \beta t^{\beta-1}
\end{equation}
For $\beta \approx 0.8$, as observed, the derivative becomes
\begin{equation}\label{ROUSEderivative3}
  \frac{d \ln(y)}{dt} \approx - \frac{\alpha \beta}{t^{0.2}}.
\end{equation}
At small $t$, this function diverges, implying that if one advances to small $t$ one has moved outside the stretched exponential's domain of validity.  At larger times, $t^{0.2}$ is nearly a constant, leading on a semilogarithmic graph to a function whose slope is nearly a constant, i.e., the function appears to be close to linear.  However, the slope does depend on time.  The apparent slope obtained from a linear fit to a section of a stretched exponential is an artifact determined by the time interval over which the fit is evaluated.

Tsalikis, et al., calculated the normalized cross-correlations $\Phi_{pqij}(0)$, eqn.\ \ref{ROUSEcrosstcfdef}, between the Rouse mode amplitudes. They observe that the cross-correlations are not large;  $|\Phi_{pqij}(0)|$ is almost always less than 0.1.  Before considering this result, we ask how large $\Phi_{pqij}(0)$ is plausibly likely to be.  If $\Phi_{pqij}(0) =1$, modes $p$ and $q$ are perfectly cross-correlated; the value of one determines the value of the other.  If one mode is cross correlated with several others, $\Phi_{pqij}(0)$ for any pair of modes must be considerably less than unity.  For example, if a given mode is equally cross-correlated with $n$ other independent modes, then at most the $n$ modes determine the value of the given mode, in which case the cross-correlations would be of typical size $1/n$.  One might also ask how accurately $\Phi_{pqij}(0)$ can be determined. If $\Phi_{pqij}(0)$ lies within simulational error of zero, non-zero values for $\Phi_{pqij}(0)$ are uninteresting.  However, the computational processes that determine $\Phi_{pqij}(0)$ and $\langle X_{pi}(t) X_{pi}(0) \rangle$ are fundamentally the same, differing only in the numbers being dropped into various computational shift registers, so the statistical error in these two quantities should be similar in size. Tsalikis, et al., followed the relaxation of $\langle X_{pi}(t) X_{pi}(0) \rangle$ through three orders of magnitude in decay, without significant scatter  appearing in the measurements, so a similar accuracy, better than one part per hundred, might reasonably be expected in determinations of $\Phi_{pqij}(0)$. Tsalikis, et al.'s, figure S.I.9 shows that measured cross-correlations $\Phi_{pqij}(0)$ can be as large as $\pm 0.1$, which does not immediately sustain the prediction of the Rouse model that $\Phi_{pqij}(0) = 0$, but perhaps a more extended examination of this obscure issue is needed.

Tsalikis, et al.,\cite{ROUSEtsalikis2017a} also evaluated the relaxation of the correlation function
\begin{equation}\label{eq:ROUSEendtoend}
   C_{uu}(t) = \langle \mathbf{u}(t) \cdot \mathbf{u}(0) \rangle.
\end{equation}
For a linear chain, $\mathbf{u}$ is the end-to-end vector.  Ring polymers have no ends, so $\mathbf{u}$ is usefully defined to be a vector from a bead to another bead half-way around the ring.  $\mathbf{u}$ has two paths to relaxation.  First, its magnitude $|\mathbf{u}|$ fluctuates around its average value, contributing a relaxation; however, this process cannot relax the correlation function to zero.  Second, as the dominant process $\langle \mathbf{u}(t) \cdot \mathbf{u}(0) \rangle$ relaxes by chain reorientation.  At long times, $\mathbf{u}(t)$ and $\mathbf{u}(0)$ cease to be correlated, so their cross-correlation function decays to zero.  As shown by the original authors, $C_{uu}(t)$ follows a stretched exponential in time, with an average relaxation time that increases as $N^{1.9}$, based on the three molecular weights studied.   Tsalikis, et al., compare $C_{uu}(t)$ from their simulations with predictions from the Rouse model.  The Rouse predictions, other than going to zero at long time, do not resemble with the simulation determinations of the time dependence of $C_{uu}(t)$.

Other quantities studied by Tsalikis, et al.\cite{ROUSEtsalikis2017a} include the intermolecular and intramolecular atom-atom radial distribution functions, which had the expected forms.   Static structure factors were calculated and found to be in good agreement with experiment. The distributions of end-to-end distances of chain segments of different lengths were calculated as functions of the length of the segments.  The distributions were in general not described by Gaussians, especially for the larger rings.  In contrast, an initial assumption of the Rouse model is that the end-to-end distances have Gaussian distributions.  Local dynamics were studied using the temporal autocorrelation functions of the torsion angles; the functions were described well with stretched exponentials in time.

Finally, these authors ask how many other polymer chains a given chain typically interacts with.  As a sensible approximation to this number, they calculated $K_{1}(r)$, the average number of other chains that had their center of mass within the radius of gyration of the chain of interest. For ring polymers $K_{1}(r)$ was in the range 1.75-2.75. For linear chains having the same three molecular weights, $K_{1}(r)$ was in the range 8.5-9.5, with $K_{1}(r)$ increasing as the chain molecular weight was increased from 5 kDa to 20 kDa.

Papadopoulos, et al.,\cite{ROUSEpapadop2016a} report united-atom simulations of polyethylene oxide rings in the melt and in dilute solution in melts of three different linear polyethylene chains. Simulations used the TrAPPE force field\cite{ROUSEfischer2008a,ROUSEfischer2008b} executed with GROMACS\cite{ROUSEabraham2015a} held at T = 413K and 1 atmosphere.  Comparison was made with simulated melts of the three linear chains and with experimental studies by Goosen, et al.,\cite{ROUSEgoossen2015a} using nuclear spin echo spectroscopy. Goosen, et al., concluded that the segmental dynamics of dilute rings in a melt of linear chains were primarily determined by the dynamics of the host polymers.   The ring polymers contained 456 monomers, for a molecular weight of 20 kDa, while the linear polymers had molecular weights of 1.8, 10, and 20 kDa, corresponding to chain $N$ of 41, 228, and 456. Simulations included 8 rings and 72-720 linear chains with $> 10^{5}$ atoms in a simulation cell.

Rouse amplitudes $X_{p}(t)$ were used to compute $\langle (X_{p}(0))^{2}\rangle$  and $\langle X_{p}(0) X_{p}(t)\rangle$, the former for $N/p^{2}$ from 100 down to $<0.02$ and the latter for $p = 2, 4, \ldots 12$.  The Rouse model predicts $\langle (X_{p}(0))^{2}\rangle \sim N/p^{2}$.  For the ring melt and the blends, this result was confirmed for $N/p^{2} >1$.  For larger $p$, i.e., $N/p^{2} < 1$, $\langle (X_{p}(0))^{2}\rangle$ deviates downward from the predicted value, attaining at the largest $p$ examined perhaps half the expected value.

\begin{figure}[thb]
\centering\includegraphics[scale=0.3]{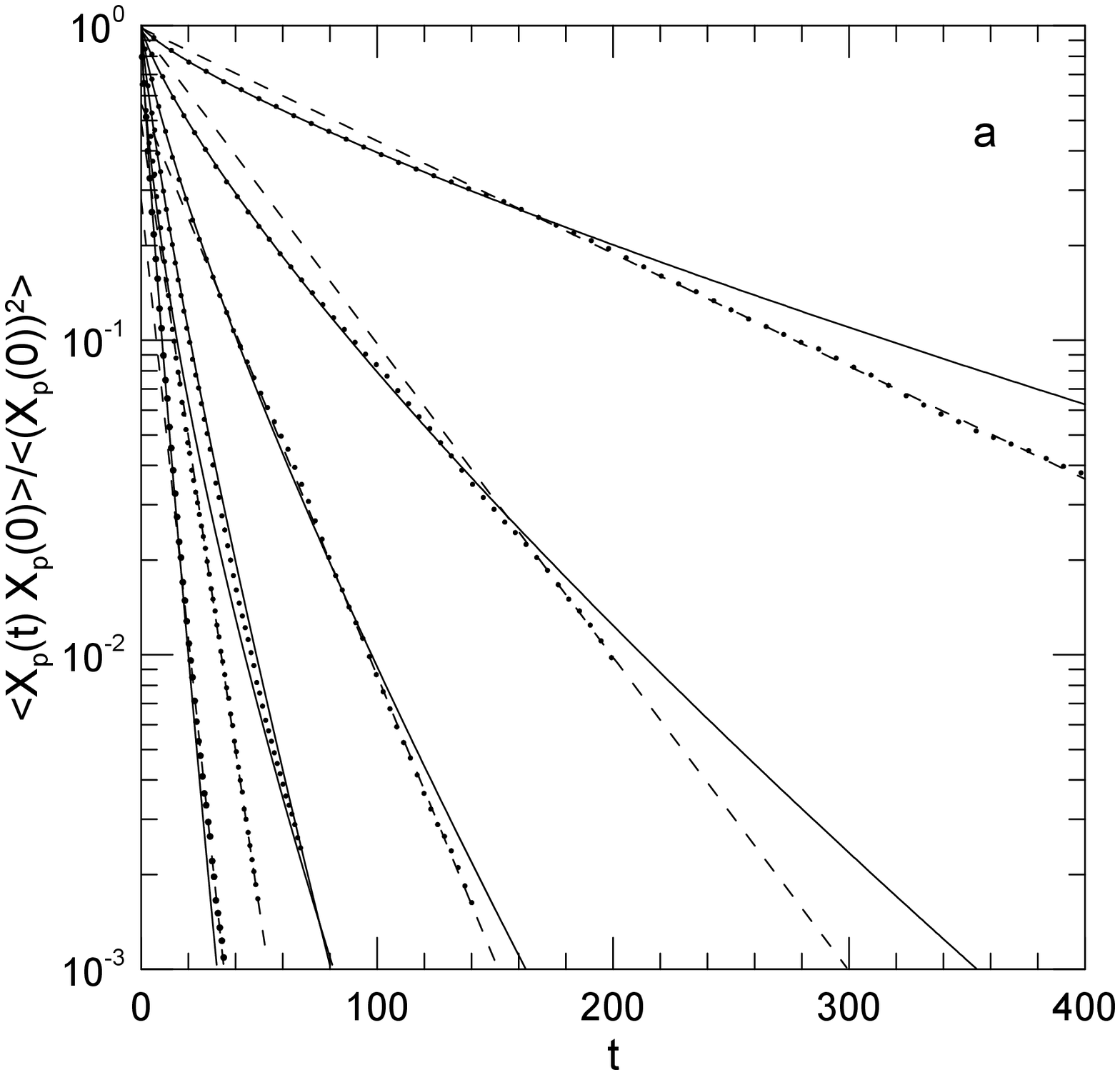} 
\centering\includegraphics[scale=0.3]{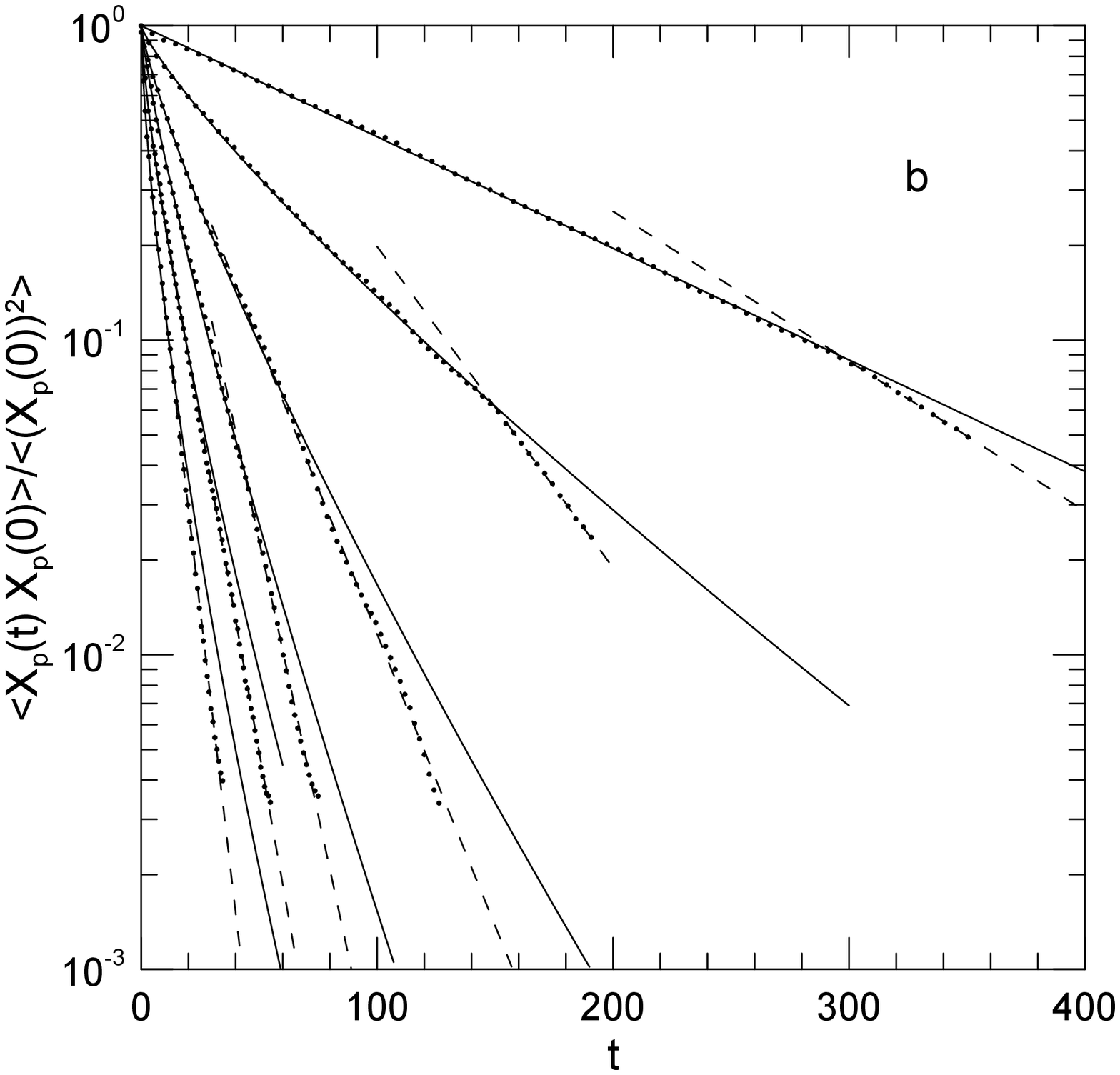}
\centering\includegraphics[scale=0.3]{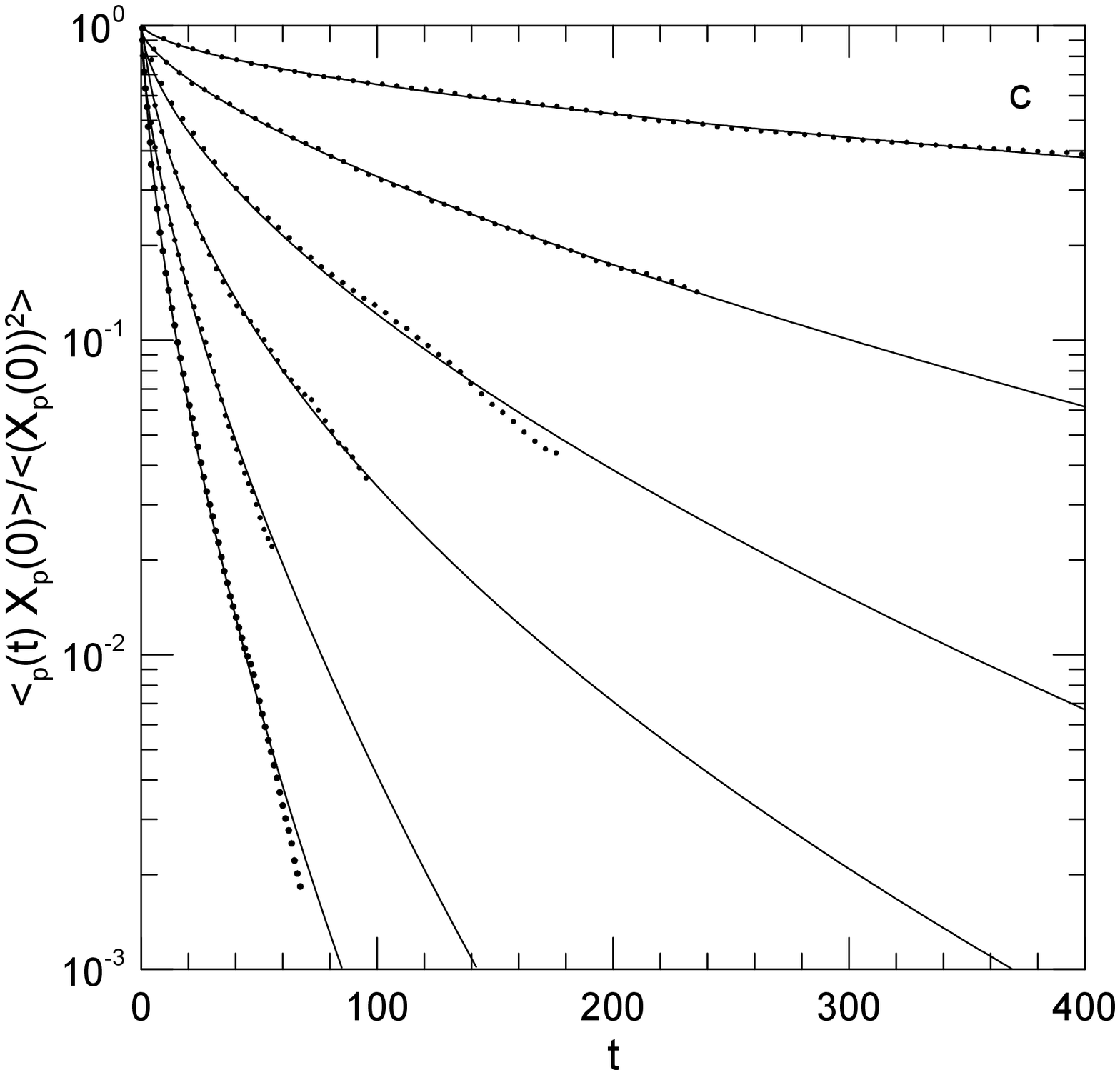}
\centering\includegraphics[scale=0.3]{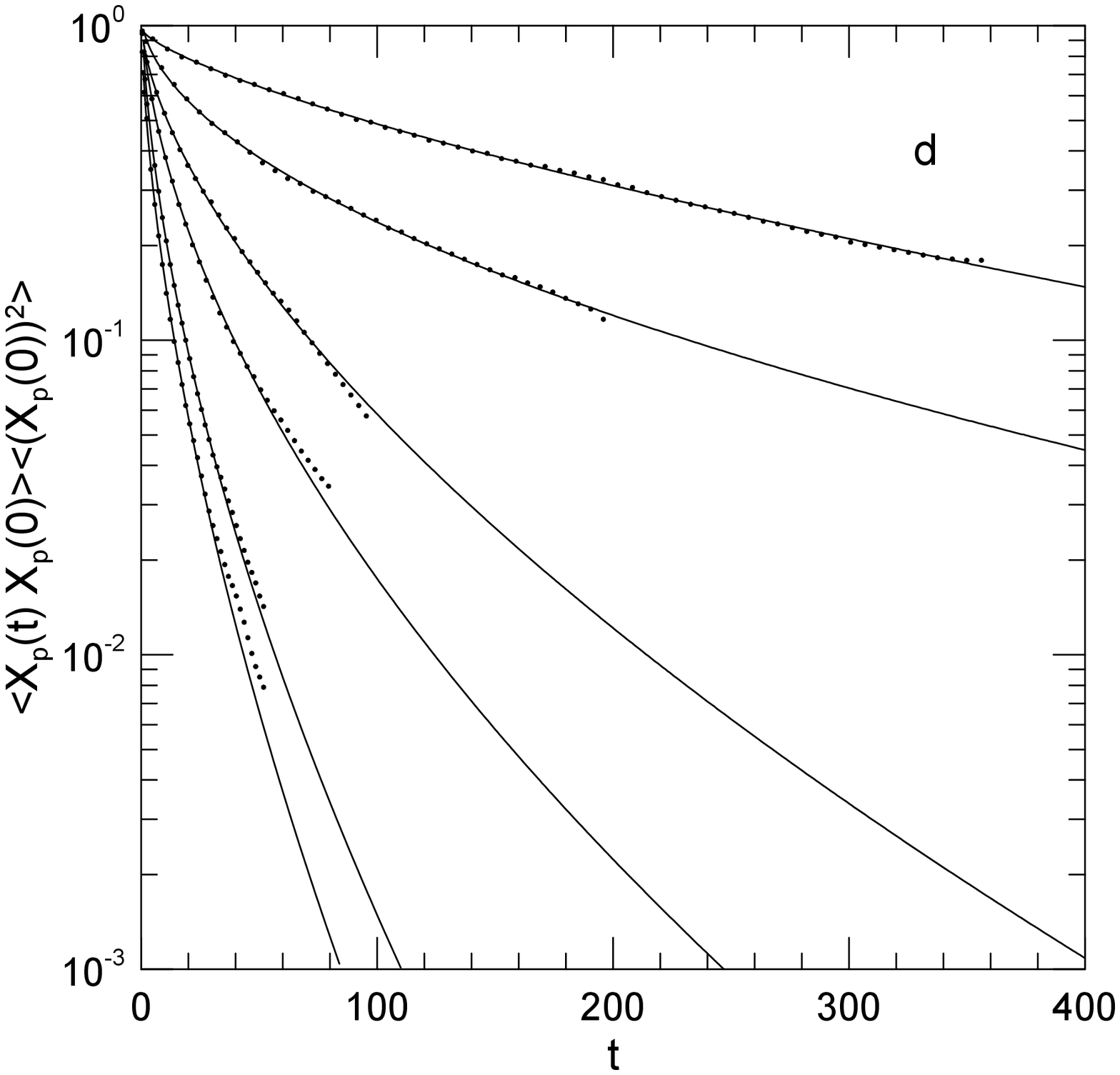}
\caption{Time autocorrelation functions $\langle X_{p}(0) X_{p}(t)\rangle$ of Rouse mode amplitudes $X_{p}(t)$ for $p =  2, 4, 6, 8, 10, 12$ (top to bottom), as normalized by $\langle (X_{p}(0))^{2}\rangle$, for (a) a 20 kDa polyethylene oxide ring melt, and dilute solutions of 20 kDa rings in melts of (b) 1.8 kDa, (c) 10 kDa, and (d)20 kDa linear polyethylene oxides.
Lines of dots are from the original simulations of Papadopoulos, et al.\cite{ROUSEpapadop2016a}, their Figure 4; solid lines represent stretched exponential fits, and dashed lines represent simple exponentials. \label{ROUSEfigpapadop2016a4}}
\end{figure}

Papadopoulos, et al.'s determinations of the time correlation functions $\langle X_{p}(0) X_{p}(t)\rangle$ appear in Figure \ref{ROUSEfigpapadop2016a4}.  They report their determinations as smooth curves, appearing in the figure here as dotted lines.  We fit to stretched exponentials (solid lines) and show simple exponentials (dashed lines) where appropriate.  There is one behavior for the ring melt and for dilute rings in the 1.8 kDa chain melt (Figs.\ \ref{ROUSEfigpapadop2016a4}$a$ and $b$), and a somewhat different behavior for dilute rings in melts of the 10 and 20 kDa chains (Figs.\ \ref{ROUSEfigpapadop2016a4}$c$ and $d$).

Our description of the relaxation functions is not entirely the same as that of Papadopolous, et al. Numerical fits clarify issues visible in the figures. In the ring melt, and in dilute solution in the 1.8 kDa linear chains, $\langle X_{p}(0) X_{p}(t)\rangle$ shows a stretched-exponential relaxation at earlier times, followed by a sharp transition to a simple-exponential relaxation at later times.  The transition, which is especially prominent for $p=2$ and $p=4$, occurs at earlier times and smaller values of $\langle X_{p}(0) X_{p}(t)\rangle/\langle (X_{p}(0))^{2} \rangle$ as $p$ is increased.  For larger $p$, the transition is more readily apparent in the solution of rings in the 1.8 kDa linear chain mely than in the ring melt.  In contrast, for rings in dilute solution in the 10 kDa and 20 kDa melts,  $\langle X_{p}(0) X_{p}(t)\rangle$ for $p=2$ and for $p=4$ relaxes as a single stretched exponential out to the longest times observed.  At larger $p$, $\langle X_{p}(0) X_{p}(t)\rangle$ fluctuates around the fitted stretched exponential.

Papadopolous, et al., report integrated times $\tau_{p}$ for their four systems and the six smallest values of $p$.  They report that $\tau_{p}$ scales approximately as $(N/p)^{2}$, $\tau_{p}$ being several-fold larger for rings in melts of the larger-$N$ linear polymers than for rings in their own melts.  Papadopoulos, et al.,'s other results are discussed in the chapter on ring polymers.  Note, however, that Papadopolous, et a., found Gaussian distributions of distances between remote parts of the rings.

\begin{table}
  \centering
\begin{tabular}{| c| c| c |c|c| c|c| c| c| c |c| }
 solvent & p & $\langle X_{p}(0)^2\rangle $ & $\alpha$ & $\beta$ & $\bar{\gamma}$  & solvent & $\langle X_{p}(0)^2\rangle $ & $\alpha$ & $\beta$ & $\bar{\gamma}$ \\  \hline
ring &   &                               &          &         &   & 1.8kDa & & &     &     \\
&2 &1&  0.025 & 0.789 & 0.008 && 1   & 0.008   & 1.003 & 0.008 \\
&4 &1& 0.063 & 0.793 & 0.027   && 1&  0.042& 0.837       &0.021    \\
&6 &1& 0.122 & 0.791 & 0.061 && 1&  0.097& 0.812       & 0.051\\
&8 &1& 0.196 & 0.812 & 0.120  && 1& 0.142&  0.830      &0.086\\
&10 &1& 0.389 & 0.654 & 0.174  && 1&  0.263&  0.739      &0.136\\
&12 &1& 0.359 & 0.852 & 0.277   && 1&  0.434&  0.679    &0.224\\  \hline
10kDa  & & & &  &  &20kDa & & & &  \\
&2 &1.015& 0.032& 0.572      &0.0015 && 0.98 &0.025& 0.720      & 0.005  \\
&4 &0.970&0.0465&0.681      & 0.0085 && 1.11 &0.135& 0.529      & 0.013\\
&6 & 0.999&0.120& 0.623      &0.023 && 1.03 &0.161& 0.627      & 0.038\\
&8 &1.28&0.326 &0.522     & 0.063  && 1.22 &0.308& 0.570      & 0.078\\
&10 &0.966& 0.269& 0.653      & 0.098 && 1.30 &0.466& 0.581      & 0.171\\
&12 &1.38&0.538& 0.584     &0.222 && 1.16&0.514& 0.590     & 0.210 \\  \hline
        \end{tabular}
  \caption{Fitting parameters for Figure \ref{ROUSEfigpapadop2016a4}.  The parameterization is $\langle X_{p}(0) X_{p}(t)\rangle = \langle (X_{p}(0))^{2}\rangle \exp(- \alpha t^{\beta})$.   \label{ROUSEpapadopoulos2016a4params}}
\end{table}

Kopf, et al.,\cite{ROUSEkopf1997a}  demonstrate a novel simulational test of the Rouse model.  They consider systems in which the forces are exactly identical, but in which the beads on some or all of the polymer chains are made four or 100 times as massive as the original 'light' beads. The forces between the beads were the FENE potential and a truncated, purely repulsive bead-bead Lennard-Jones potential.  From basic statistical mechanics, this change should have no effect on the static properties of the chains in a melt, an outcome that was confirmed simulationally.  In mixtures, increasing the mass of the heavier beads slows down the motions of the light beads.  Kopf, et al., took advantage of the fact that they were doing molecular dynamics to calculate the velocity autocorrelation function through multiple oscillations out to long times. The frequency of the oscillations is relatively independent of the fraction of light or heavy polymers in the system, suggesting that the oscillations in the velocity autocorrelation function arise primarily from intramolecular interactions. The Rouse model remained accurate in light-heavy polymer mixtures.  Rouse modes were found not to be cross-correlated. Rouse amplitude autocorrelation functions decayed approximately exponentially in time.  Contrary to the Rouse model, these simulations observed subdiffusion (mean-square center-of-mass displacement proportional to $t^{0.8}$) on shorter time scales.  A nominal entanglement time was used to estimate a nominal tube diameter, which the authors also described as a characteristic length for slowing down of monomer motion. Their results indicated that the tube nominal diameter is independent of the monomer mass, implying that the tube diameter is a static rather than a dynamic quantity, consistent with topological pictures for entanglements.

We turn to Paul, et al.,\cite{ROUSEpaul1997a} who studied a C$_{100}$  polyethylene using atomistic molecular dynamics.  Their simulations included both an explicit-atom model and also a unified atom model in which each CH$_{2}$ group was treated as a single atom.  The polymer was chosen to be long enough that it could reasonably be expected to show Gaussian behavior for its static chain statistics, yet short enough that its dynamics would be expected to have Rouse-like and not reptational behavior.  The authors recognized that the assumption of Rouse-like behavior in unentangled melts required examination. A stochastic dynamics simulation was used to equilibrate the samples, while data was obtained using molecular dynamics.  Static behavior was tested by calculating the static structure factor; good agreement between simulation and experiment was found. In addition to other dynamic studies, large-scale dynamic behavior was compared with expectations from the Rouse model.

The end-to-end vector reorientation time and the long-time self diffusion coefficient are consistent with the same value for the segmental friction coefficient, these results being applicable '\emph{on time scales larger than the Rouse time.}' However, contrary to the Rouse model, at times shorter than the Rouse time the center-of-mass diffusion is subdiffusive, being proportional to $t^{0.83}$ or so.  Static mean-square amplitudes $\langle (X_{p}(0))^{2}\rangle$ of Rouse modes were calculated.  For $p \leq 3$, the Rouse model expectation $\langle (X_{p}(0))^{2}\rangle \sim p^{-2}$ was observed. For $p >3$, the mean-square mode amplitudes decrease approximately as $p^{-3}$, not the $p^{-2}$ predicted by the Rouse model. The equal-time cross-correlation functions $\langle X_{p}(0) X_{q}(0) \rangle$ ($p \neq q$) were found to vanish to ``\emph{...within the error bars in the simulation}.''

Paul, et al., also calculated the temporal autocorrelation functions $\langle X_{p}(t) \cdot X_{p}(0)\rangle$.  A plot of correlation functions with  $p = 1, 2,$ and $3$  finds that the three correlation functions decay nearly exponentially as $\exp(- \Gamma p^{2} t)$, a single value of $\Gamma$ sufficing for all three values of $p$, with small deviations over the first quarter of the decay.  For $p>3$, the $\langle X_{p}(t) \cdot X_{p}(0)\rangle$ are markedly non-exponential.  When plotted against $p^{2}t$, with increasing $p$ the $\langle X_{p}(t) \cdot X_{p}(0)\rangle$  decay more rapidly. The authors conclude that the Rouse model '\ldots \emph{is at most applicable to a few largest scale eigenmodes.}'  They do, however, note that the self-diffusion coefficient and the rotational diffusion coefficient can be described self-consistently in terms of a single segmental friction factor.

These results were extended by Paul, et al.,\cite{ROUSEpaul1998a} who considered the single-chain intermediate structure factor $g^{(1)}(q,t)$ for unentangled polyethylene molecules in a melt, comparing results from neutron spin echo spectroscopy with results from atomistic and from united-atom molecular dynamics simulations. They continued to study C$_{100}$ polyethylene, because the polymer is short enough not to be entangled and long enough to have Gaussian chain statistics.  The corresponding Rouse model has two parameters, namely a bond strength revealed by the average segment length $\sigma$, and a monomer drag coefficient revealed by the chain center-of-mess diffusion coefficient $D$, the latter determined both experimentally and from each of the two sets of simulations.  The simulation values for $g^{(1)}(q,t)$ for the explicit-atom and unified-atom simulations were in agreement with the experiments over a factor of 6 in $q$ and two orders of magnitude in the scaled time $Dt$.

Having validated the accuracy of the simulations, Paul, et al., then used the simulations to calculate the Rouse amplitudes, their time autocorrelation functions, and the $g^{(1)}(q,t)$ implied by the Rouse modes.   The $g^{(1)}(q,t)$ predicted by the Rouse model only agrees with the simulations for a limited range of $q$ ($\leq 0.14 \mbox{\AA}^{-1}$) and times $\leq 4$nS. At larger $q$ and at also longer times, Rouse-model predictions of $g^{(1)}(q,t)$ are significantly smaller than $g^{(1)}(q,t)$ from experiment or as calculated directly in the simulations. The authors note three marked deviations between the simulational results and the Rouse model: First, for times $\leq \tau_{R}$, diffusion is found by the simulations to be subdiffusive, with exponent 0.83, rather than diffusive; in contrast, Rouse-model chains always exhibit normal center-of-mass diffusion.  Second, simulations find that only the lowest Rouse modes $p \leq 3$, have relaxations that scale as $p^{2} t$; in the Rouse model, all modes have this property.  Third, in the simulations each $\langle X_{p}(t) \cdot X_{p}(0)\rangle$ decays as a stretched exponential in time; the $\langle X_{p}(t) \cdot X_{p}(0)\rangle$ of the Rouse model are all pure exponentials.

Finally, $g^{(1)}(q,t)$ from the simulations, together with the Gaussian approximation
\begin{equation}
  \label{eq:ROUSEgaussianapproximation}
  g^{(1)}(q,t) = \exp(- q^{2} \langle (\Delta R_{\rm cm}(t))^{2} \rangle /6),
\end{equation}
was used to calculate a mean-square displacement $\langle (\Delta R_{\rm cm}(t))^{2} \rangle$. It should again be emphasized that Doob's theorem guarantees as a mathematical certainty that if the physical requirements leading to the Gaussian approximation are valid, then as a mathematical certainty $\langle (\Delta R_{\rm cm}(t))^{2} \rangle$ increases linearly in time.   However, as found by Paul\cite{ROUSEpaul1997a} at short times the calculated center-of-mass motion is subdiffusive, i.e., $\langle (\Delta R_{\rm cm}(t))^{2} \rangle$ grows as $t^{0.83}$ not as $t^{1}$.  The Gaussian-approximation estimate of the mean-square displacement agrees with the simulations at long times $t \geq \tau_{R}$, at which the center-of-mass motion is diffusive.   At times shorter than the Rouse time, $\langle (\Delta R_{cm}(t))^{2} \rangle$ as determined by the simulation is considerably larger than $ \langle (\Delta R_{cm}(t))^{2} \rangle$ inferred from $g^{(1)}(q,t)$ and equation \ref{eq:ROUSEgaussianapproximation}, showing that the Gaussian approximation is not valid in these systems at shorter times.

Several theoretical advances followed this work. Smith and Paul\cite{ROUSEsmith1998b} used quantum chemistry calculations to generate an improved set of force parameters for simulations of 1,4-polybutadiene. Harnau, et al.\cite{ROUSEharnau1999a,ROUSEharnau1999b} proposed that these results of Paul, et al.\cite{ROUSEpaul1998a} could be understood by replacing the Rouse model with a semiflexible chain model that takes into account chain stiffness.  The semiflexible chain model with reasonable parameters agrees well with Paul, et al.'s experimental and simulational determinations of $S(q,t)$.

Smith, et al.,\cite{ROUSEsmith2001b} present simulations of an unentangled polybutadiene melt, the focus of the work being to examine the presence of non-Gaussian displacement distributions of polymer beads in a melt. The single-chain intermediate structure factor $g^{(1)}(q,t)$ was determined from neutron spin-echo measurements and separately from molecular dynamics simulations using Smith and Paul's\cite{ROUSEsmith1998b}united-atom potential. For $0.05 \leq q \leq 0.30$ \AA$^{-1}$ and times out to 17 nS, measured and simulated values of $g^{(1)}(q,t)$ were in good agreement.  The center-of-mass motion was diffusive at long times but subdiffusive ($\langle (\delta R(t))^{2} \rangle \sim t^{0.8}$) at times shorter than $\tau_{R} \approx 15$ nS. The simulated $g^{(1)}(q,t)$ was compared with predictions of the Rouse model and several of this model's proposed modifications, finding that none of the models reproduced the simulations.  Simulations were also use to calculate $\langle X_{p}(t)X_{q}(0)\rangle$, the correlation function vanishing for $p \neq q$, at least for $p, q \leq 4$. Use of the simulated $\langle X_{p}(t)X_{q}(0)\rangle$ in the Rouse form for $g^{(1)}(q,t)$ also did not lead to agreement of this modified Rouse model with experiment.  The authors note that the Gaussian approximation for $g^{(1)}(q,t)$ is only appropriate if the distribution of bead displacements $\mathbf{R}_{m}(t)-\mathbf{R}_{n}(0)$, $\mathbf{R}_{m}(t)$ being the position of bead $m$ at time $t$, is Gaussian at all times. To examine the consequences of this observation, they calculated $g^{(1)}(q,t)$ using the Gaussian approximation and values of mean-square displacements  $\langle(\mathbf{R}_{m}(t)-\mathbf{R}_{n}(0))^{2}\rangle$, finding that this calculated $g^{(1)}(q,t)$ was in good agreement with $g^{(1)}(q,t)$ as predicted by the Rouse model, but did not agree with $g^{(1)}(q,t)$ as calculated directly from the simulation, thus showing the importance of non-Gaussian particle displacements.  Smith, et al., conclude that the non-Gaussian distribution of bead displacements $\mathbf{R}_{m}(t)-\mathbf{R}_{n}(0)$ is responsible for the observed failure of the Rouse model in polymer melts. Two sorts of non-Gaussian behavior possible here.  The first is that the distribution of displacements $P(\Delta \mathbf{R}_{m}(\Delta t))$ for each bead separately could be non-Gaussian. The second is that the distributions of displacements of pairs of beads could be cross-correlated.  Thanks to the fluctuation-dissipation theorem, this latter possibility is equivalent to the statement that there are significant hydrodynamic interactions in polymer melts, a possibility that would only be surprising if the Rouse model were correct in polymer melts.

Harmandaris, et al.,\cite{ROUSEharmandaris1998a}, made atomistic simulations of 24-, 78-, and 156-atom (mean length) linear polyethylene melts, finding a diffusion coefficient $D$ as well as the time autocorrelation functions of the polymer end-to-end vector and the Rouse mode amplitudes. The study was novel in that the authors deliberately simulated polydisperse melts having polydispersity index near 1.09.  Each autocorrelation function may be said to have a characteristic time $\tau$.  From these quantities, nominal monomer friction factors were extracted.  Initial chain configurations were equilibrated using the end-bridging Monte Carlo scheme\cite{ROUSEpant1995a}. Molecular dynamics were executed using a sixth-order predictor-corrector model.  The objectives of the study were to test the Rouse model, and to take advantage of the polydispersity to examine the dynamics of chains having different lengths, all in the same melt. Potential energies included a Lennard-Jones potential between non-bonded atoms, bond-bending and torsional potentials, and a Fixman potential\cite{ROUSEfixman1974a} to keep bond lengths constant. The 24- and 78-atom carbon models were simulated in both the NVE and NVT ensembles; results agreed. The mean-square end-to-end distance $\langle (\mathbf{r}(0))^{2}\rangle$, radius of gyration, and intermolecular bead-bead static correlation functions from the molecular dynamics simulation and the end-bridging Monte Carlo simulation were found to agree, confirming the validity of the two simulations. Local dynamics as estimated with the torsion angle temporal autocorrelation function showed that local dynamics become slightly slower as the chain length is increased.

Harmandaris, et al.,\cite{ROUSEharmandaris1998a} calculated properties of the Rouse amplitudes $X_{p}(t)$.  The mean-square static amplitudes $\langle (X_{p}(0))^{2} \rangle$ decrease with increasing $p$, much more rapidly than the $p^{-2}$ dependence predicted by the Rouse model.  For $p=10$, the discrepancy between the simulation and the Rouse model approaches an order of magnitude.  The temporal correlation functions $\langle X_{p}(0) X_{p}(t) \rangle$, at least for the 83- and 117-carbon chain  systems, also do not agree with the Rouse model, namely they are not simple exponentials, and their relaxation times do not scale with time as $p^{2}t$. On the other hand, for the end-to-end vector, the calculated $\langle \mathbf{r}(0)\cdot \mathbf{r}(t)\rangle$ using the Rouse model and a relaxation time inferred from  $\langle X_{1}(0) X_{1}(t) \rangle$ agrees well with $\langle \mathbf{r}(0)\cdot \mathbf{r}(t)\rangle$ obtained from the simulation.  From observations of the chain center-of-mass motion over long times, a chain diffusion coefficient $D$ and therefore a monomer friction factor $\zeta$ can be inferred.  Contrary to the Rouse model, $\zeta$ depends on chain length, increasing threefold from the shortest to the longest chains studied.  However, it is plausible that $\zeta$ is reaching an asymptotic value for the longer chains.  Harmandaris, et al., also calculated, from the diffusion coefficient, a zero-shear viscosity.  The calculation was based on Rouse's theory, which in most other respects does not describe the dynamics of these systems.

Krushev, et al.,\cite{ROUSEkrushev2002a} simulated melts of 1,4-polybutadiene.  Their interest was to determine the effects of torsion barriers on molecular motions.  To do this, they examined the static structure factor, Rouse mode amplitudes and Rouse-Rouse temporal autocorrelation functions, and the intermediate scattering function $S(q,t)$.  Their polymer melts incorporated 40 polymer chains, each with 29 or 30 subunits, all with united atom potentials, including a model with chains incorporating vinyl groups, a model with chains not incorporating vinyl groups, and a model with no vinyl groups and all rotational potentials set to zero. The three models have the same distribution for their radii of gyration.  Rouse mode amplitudes had at most weak cross-correlations, $\langle X_{p}(0) X_{q}(0)\rangle$ for $p \neq q$ being less than 2\% of  $\langle (X_{p}(0))^{2} \rangle$.  The mode amplitudes $\langle (X_{p}(0))^{2} \rangle$ did not, however, follow the Rouse prediction $\langle (X_{p}(0))^{2} \rangle \sim p^{-2}$; the mean-square amplitude instead was identified as following a prediction for a freely rotating polymer.\cite{ROUSEkreer2001a}  The calculated static structure factor was not affected by adding or deleting the torsion potential.  $\langle X_{p}(t) X_{p}(0)\rangle$ showed stretched-, not single-exponential behavior for all $p$ studied over an adequate time range, time correlation function with and without torsion potentials having being very nearly the same when plotted in reduced time units in which the $1/e$ time was unity.  The intermediate scattering function $g^{(1)}(q,t)$ also did not follow the Rouse model predictions, namely the Rouse model predicts an over-rapid decay of $g^{(1)}(q,t)$ at larger $q$, and underpredicts the degree of stretching of $g^{(1)}(q,t)$.  $g^{(1)}(q,t)$ is not significantly changed when torsional potentials are added or removed from the potential energy; the authors infer that observed deviations from Rouse behavior are not due to internal rotation barriers.  The Rouse model calculation of $g^{(1)}(q,t)$ agrees with the $g^{(1)}(q,t)$ calculated from the simulation on using the Gaussian approximation.  Neither calculation agrees with the actual $S(q,t)$.  Furthermore, the center of mass mean-square displacements are subdiffusive at short times. \emph{The authors conclude that the Rouse model assumption that atomic motions are described by a joint Gaussian random process is thus shown to be incorrect.}

Padding and Briels\cite{ROUSEpadding2001b} report simulations of a C$_{120}$H$_{242}$ polyethylene melt. They simulated four different starting points for their melt, molecules having united-atom potential energies.  Bond lengths and angles had harmonic potential energies; a torsion potential energy was present; unbonded (separated by four or more atoms in a single molecule) pairs of united atoms have a Lennard-Jones potential.  The potential energy used by these authors is not the simple-harmonic bond-length potential of the Rouse model. A weak coupling to a bath held the temperature fixed. Melt starting chain configurations were created by gradual compression of a dilute system in which only repulsive interatomic forces were present.

The time-dependent dynamic parameters that they obtained from their simulations include mean-square displacements, the end-to-end vector time autocorrelation function, the dynamic structure factor, and the stress tensor. A single set of numerical parameters for the number of segments $N$ in a chain, for the diffusion coefficient $D$, and for the longest relaxation time $\tau_{1}$ described most of the dynamic quantities that they calculated, but only over distances longer than a limiting length scale,.  This paper actually tested the relationships between $N$, $D$, $\tau_{1}$ and the calculated dynamic parameters, as predicted by the Rouse model, but did not test the Rouse model itself, Rouse's description of the internal dynamics of a polymer chain.

Here the stress tensor was calculated as
\begin{equation}
      \label{eq:ROUSEstresstensor}
      \mathbf{\sigma}  = \frac{1}{V} \sum_{i=1}^{N} m_{i} \mathbf{v}_{i}\mathbf{v}_{i} + \sum_{i \geq j = 1}^{N}  (\mathbf{r}_{i}-\mathbf{r}_{j}) \mathbf{F}_{ij}
\end{equation}
where $\mathbf{v}_{i}$ is the velocity of atom (or center of mass) $i$, $\mathbf{r}_{i}$ is the position of atom (or center of mass) $i$, and $\mathbf{F}_{ij}$ is the force exerted on atom (molecular center of mass) $i$ by atom (molecular center of mass) $j$.  The forces between two molecules can create a torque, an antisymmetric part of the stress tensor, on each molecule.  The stress tensor was identified as leading to the zero-shear relaxation modulus via
\begin{equation}
   \label{eq:ROUSErelaxationmodulus}
   G(t) = \frac{V}{10 k_{B} T} \langle \mathbf{P}(t):\mathbf{P}(0) \rangle,
\end{equation}
where $\mathbf{P}$ is the symmetrized traceless part of $\mathbf{\sigma}$.

Padding and Briels concluded that there is a shortest length scale $b \approx 1.2$ nm over which the Rouse model is valid in their simulations.  The length scale manifests itself as the shortest Rouse-mode wavelength for which the model works, the shortest distance over which a mean-square displacement must occur before Rouse behavior is seen, and the shortest wavelength for which the simulated $S(q,t)$ agrees with the model.  Over shorter distances and times, matters became more complicated. Padding and Briels calculate $S(q,t)$ for a series of wave vector magnitudes.  At small $q$, only center-of mass diffusion is seen.  At larger $q$, $S(q,t)$ has contributions from polymer internal modes.  At larger $q$, $S(q,t)$ from the simulation and $S(q,t)$ calculated from the Rouse modes and the diffusion coefficient found at small $q$ do not agree, $S(q,t)$ from the Rouse formula decaying faster at long times than $S(q,t)$ from the simulations, the discrepancy becoming larger at larger $q$.  The short-distance internal chain modes are thus not the same as the internal modes predicted from the Rouse model.  Similarly, measurements of mean-square displacements from simulations only agree with the Rouse model when mean-square displacements are greater than (1.1 nm)$^{2}$.  Finally, at short times the simulated shear relaxation modulus $G(t)$ ``...does not behave Rouse-like at all...'', but corrections due to this issue at long times were limited in size.

In a further paper, Padding and Briels\cite{ROUSEpadding2001c} report simulations of a heavily coarse-grained (one bead = 20 monomers, 120 chains in a simulation box) C$_{120}$ linear polyethylene that incorporated a complicated switchable scheme for enforcing chain uncrossability.  The scheme could be turned off, leading to simulations of a melt in which polymer chains could pass through each other.  Interactions included non-bonded, bonded, and bending-angle contributions to the bead-bead potential of average force.  Beads incorporated as many monomers as feasible without making the beads larger than a nominal tube radius.  Padding and Briels calculated the mean-square displacements, both of single blobs and of the chain centers of mass.  For the unentangled chains, mean-square displacements increased linearly with time.  Adding the uncrossability constraint reduced the mean-square displacements and gave them a sublinear time dependence over a considerable time regime.

Padding and Briels\cite{ROUSEpadding2001c} calculated the time-dependent Rouse amplitudes and evaluated their time correlation functions. Fits were then made to stretched exponentials in time.   Chains had six beads, so they only had five internal Rouse modes. When chain crossing was permitted, the time correlation functions were very nearly single exponentials.  Adding chain crossing constraints and a bond bending potential led to appreciably non-exponential relaxations, the stretching parameter falling from close to unity in the absence of chain crossing constraints or a bond bending potential to 0.77 for the three highest modes when the constraint and potential were added.  The chain crossing constraint considerably increased the relaxation times of the $p = 1$ and $p=2$ modes but did not increase substantially the relaxation times of the three higher-$p$ modes.  Padding and Briels determined effective relaxation times $\tau^{\rm eff}_{p}$ for their five modes.  However, instead of calculating $\tau_{p}$ from the stretched-exponential fitting parameters, the authors did numerical integrals of the measured $\langle X_{p}(t) X_{p}(0) \rangle$ curves.  The Rouse relaxation rates, equation \ref{eq:ROUSEweffective} were evaluated for each $p$.  In the Rouse model, $W^{\rm eff}_{p}$ is independent of $p$.  In the presence of chain stiffness, and more dramatically in the presence of uncrossability, $W^{\rm eff}_{p}$ was found to increase several-fold as $p$ was increased, the major change from the Rouse model being that in the presence of chain stiffness and uncrossability $W^{\rm eff}_{p}$ is reduced for small $p$.  Finally, Padding and Briels\cite{ROUSEpadding2001c} calculated the system's dynamic structure factor $S(q,t)$ for a series of values of $q$.  At small $q$, $S(q,t)$ is relaxed by whole-body translational diffusion, a fit giving the polymer's diffusion coefficient.  In the presence of chain uncrossability and larger $q$, $S(q,t)$ did not agree with the Rouse model predictions.

Padding and Briels\cite{ROUSEpadding2002a} further extended their work on polyethylene by making extended united-atom simulations of melts of seven different polyethylenes, using chains with 80 to 1000 carbon atoms coarse-grained into 4 to 50 blobs at 450 K and a density 0.761 g/cm$^{3}$.  They stress that the eliminated internal coordinates become thermal bath variables, and contribute to the motion of the blobs as unseen random thermal forces and a friction factor, which they treated as a scalar with no associated memory function. The paper considers a considerable list of different dynamic parameters; this chapter is only concerned with the behavior of the Rouse amplitudes.  They calculated the Rouse amplitudes and their time correlation functions.  On plotting $ \langle (X_{p}(0))^{2}\rangle 8 N \sin^{2}(p\pi/2N)$ against $N/p$, values of  $\langle X_{p}(0)^{2}\rangle$ for all chain lengths superpose, but contrary to the Rouse model $\langle X_{p}(0)^{2}\rangle$ is not independent of $N/p$; it instead falls off from slightly more than 3 to slightly more than 1 as $N/p$ is reduced, i.e., as $p$ is increased..

Padding and Briels\cite{ROUSEpadding2002a} also calculated the Rouse-Rouse time correlation functions.  Rouse modes at time zero are uncorrelated; they did not report what happens at later times.
Rouse temporal autocorrelation functions were found to decay as stretched exponentials in time.  The stretching parameter $\beta$ was close to 0.7 near $N/p = 1$.  $\beta$ decreased to 0.55 or so near $N/p =2$, and then increased to near 0.7 at $N/p=6$.  At larger $N/p$, $\beta$ was nearly constant.  The dependence of $W^{\rm eff}_{p}$ on $N/p$ was examined.  For $N/p \approx 1$, $W^{\rm eff}_{p}$ is independent of $N/p$. For $N/p$ modestly above 1 and out to 3 or so, $W^{\rm eff}_{p} \sim (N/p)^{-2}$.  For $N/p > 3$, $W^{\rm eff}_{p} \sim (N/p)^{-1}$.  Padding and Briels note that for the second and third regimes these dependences are not in agreement with either the Rouse or the reptation model.  Padding and Briels then propose that at large times $\langle X_{p}(t)X_{p}(0)\rangle$ switches over from a stretched-exponential to a simple-exponential time dependence.

Abrams and Kremer\cite{ROUSEabrams2002a} studied a bead and spring polymer melt, the interest being the effects of varying the equilibrium bond length $\ell_{0}$ relative to a nominal bead diameter $d_{0}$. In different simulations, the bead length was given 13 values in the range $0.73 \leq \ell_{0}/d_{0} \leq 1.34$.  The model contained 80 freely-jointed chains, each having 50 beads, at density 0.85 $\sigma^{-3}$ and nominal temperature $T=1$ in natural units, so that $d_{0} = \sigma$.  Bonded beads were linked with a harmonic potential $\frac{k}{2}(\ell-\ell_{0})^{2}$, $\ell$ being a bond length, with $k$ and $\ell_{0}$ being simulational parameters.  Non-bonded beads interacted with a truncated Lennard-Jones potential.  The authors studied the time correlation functions of the Rouse amplitudes $ X_{p}(t)$ and the mean-square displacements of the bead centers-of-mass.

Semi-log plots of the normalized $\langle X_{p}(t) X_{p}(0)\rangle$ as functions of a normalized time $t p^{2}/N^{2}$ were presented for $\ell_{0}/d_{0}$ equaling 0.79, 0.97, and 1.24 and $p \leq 5$.   For $\ell_{0}/d_{0} = 0.79$ , the plots were nearly linear.  The curvature increased with increasing $\ell_{0}/d_{0}$.  For the smallest $\ell_{0}/d_{0}$, plots of $\langle X_{p}(t) X_{p}(0)\rangle$ for the different values of $p$ nearly superpose. For larger $\ell_{0}/d_{0}$,the curves spread out modestly from each other, though the dependence on $p$ is hard to discern.  Abrams and Kremer extracted $\tau_{p}$ from fits of $\exp(-t/\tau_{p}$ to the early parts of the $\langle X_{p}(t) X_{p}(0)\rangle$ curves, and advanced from there to nominal friction factors $\zeta_{p}$, invoking an assumption that the Rouse model was adequately valid at earlier times.  The inferred $\zeta_{p}$ values were presented as averages over $p$.  As a function of $\ell_{0}/d_{0}$, the averaged $\zeta_{p}$ increase rapidly at larger $\ell_{0}/d_{0}$. Abrams and Kremer also calculated the average number of other polymer beads within a distance $2^{1/6} \sigma$ of a bead of interest.  That number increases, roughly from 0.4 to 1, over the observed range of $\ell_{0}/d_{0}$.

Doxastakis, et al.\cite{ROUSEdoxastakis2003a} report extensive atomistic and unified-atom simulations of very short (40-115 atom) polyisoprenes, and compare with measurements from $^{13}$C NMR, quasielastic neutron scattering, the torsional correlation function from the simulation, dielectric relaxation spectroscopy, and polymer self-diffusion.  Because these authors did atomistic simulations, their simulations determined single-bond and few-atom motions that could be compared with $^{13}$C NMR and neutron scattering. Reasonable agreement between simulation and experimentally measured quantities, within the expected limits of accuracy of the simulations, was obtained.  Dielectric relaxation measurements were interpreted in terms of a Kohlrausch-Williams-Watts function for a higher frequency peak and Rouse normal modes for a lower-frequency peak. The Rouse fit showed some deviation from experiment at higher frequencies. The mean-square amplitudes $\langle (X_{p}(0))^{2} \rangle$ of the Rouse modes only followed the theoretical $p^{-2}$ scaling for the first two or three modes; for larger $p$ the measured amplitudes are smaller than the theoretical prediction. Plots of the simulated $\langle X_{p}(0) X_{p}(t) \rangle$ against $p^{2} t$ should collapse onto a single line.  If the $t=0$ amplitude is normalized out, the plots come respectably close to doing so.  However, at short times  $\langle X_{p}(0) X_{p}(t) \rangle$ from the simulations fell well below a fit of the long-time $\langle X_{p}(0) X_{p}(t) \rangle$ to a single exponential, especially for larger $p$.  The simulated time autocorrelation function for the chain end-to-end vector is at early times also smaller than expected from the Rouse model.  Finally, on uniting the various theoretical and fitted treatments of chain end-to-end relaxation, very good agreement is obtained between the theoretical form and the simulations. The authors conclude that the Rouse model is sustained by their simulations, for the quantities that they analyzed, a conclusion that neglects the issues they faithfully reported with the mode mean-square amplitudes.

Tsolou, et al.\cite{ROUSEtsolou2005a,ROUSEtsolou2008a,ROUSEtsolou2010a} report a series of molecular dynamics simulations of polybutadiene and polyethylene.  Their first paper\cite{ROUSEtsolou2005a} simulated \emph{cis}-1,4-polybutadiene based on a united atom description in which hydrogen atoms were merged with the carbon atom to which they were bonded.  Bonds were represented as Hookian springs of finite rest length; bend and torsion angles had associated potential energies. Non-bonded atoms interacted with a non-truncated Lennard-Jones potential. Melt simulations were done on monodisperse polymers having $N$ of 32 to 400 carbon atoms for times out to 600 nS. End-bridging Monte Carlo methods were used to create rapid equilibration; simulations were based on multiple-time-step molecular dynamics. The system was thermostatted to constant temperature and pressure. A long series of static quantities were calculated, including the mean-square radius of gyration, mean-square end-to-end distance, characteristic ratio, specific volume as a function of chain length, density as a function of temperature, the intermolecular pair radial distribution function, and the static structure factor.  For the last of these, the locations of the hydrogen atoms had to be backed out from the unified atom description.

Tsolou, et al.,\cite{ROUSEtsolou2005a} also calculated dynamic quantities, including the time autocorrelation functions for the torsion angles and the chain end-to-end vector $\mathbf{R}(t)$.  Efforts to fit $\langle \mathbf{R}(t) \cdot \mathbf{R}(0) \rangle$ as a sum of Rouse modes were unsatisfactory; on the other hand,  $\langle \mathbf{R}(t) \cdot \mathbf{R}(0) \rangle$ was fit accurately with a single stretched exponential in $t$. The nominal relaxation times from these fits increased with increasing  polymer length $N$, namely $\tau \sim N^{a}$ with $a$ increasing from 2.1 for the shortest polymers to 2.8 for the longest polymers.  The change in $a$ with increasing $N$ was not obviously discontinuous.

Tsolou, et al., also determined the mean-square displacements of the chain centers-of-mass as functions of time, and inferred from these dependences the diffusion coefficient $D$.  $D$ depends on $N$ approximately as a power law.  Curiously, the slope of $\log(D)$ against $\log(N)$ is shallowest for intermediate values of $N$.

An algorithm was used to obtain a nominal primitive path for each polymer chain at a series of times.  The primitive path from the algorithm is a smooth curve that follows the atomistic backbone. The diameter $a$ of the corresponding tube is $64\text{\AA }$, which is considerably larger than the experimental $38\text{\AA }$ tube diameter reported\cite{ROUSEfetters1994a} for the same system.  $a$ is  much larger than the distance between neighboring polymer chains, showing that when a polymer chain attempts to move transversely to its primitive path, and encounters another polymer chain, in general it is able to continue to move in the same direction over considerable distances.  The authors also computed the mean-square displacements $g(t)$ of the central beads of each chain.  $g(t)$ appears to be a smooth curve that could be described as having sections that follow power-laws $t^{\alpha}$.   However, $\alpha$ was never less than 0.4, and never reached the 0.25 of the Rouse model. For $N=400$, the transition from an initial $\alpha = 0.5$ down to $\alpha = 0.4$ occurred at $\approx 10^{3}$ pS, i.e., several nS.

Calculations of the single-chain dynamic structure factor $g^{(1)}(q,t)$ were made.  The authors concluded that is no $q$ for which $g^{(1)}(q,t)$ agrees with a fit to the Rouse model, including times shorter than a few nS at which, according to the tube-reptation model, polymer chains are supposed to be performing Rouse-like motion, because they have not yet having encountered the walls of their tube.  $g^{(1)}(q,t)$ obtained from these simulations decays more slowly than does $g^{(1)}(q,t)$ predicted by the Rouse model, using Rouse times calculated by fitting independently to the time correlation functions of the polymer end-to-end vector.  Nonetheless, the authors were able to extract a nominal friction factor $\zeta$ from the Rouse form for the diffusion coefficient, even though the Rouse model does not appear to describe the dynamics.

Tsolou, et al.,\cite{ROUSEtsolou2008a} examined Rouse amplitudes and dynamic structure factors for simulated cis-1,4-polybutadiene melts.  The simulations viewed 32 chains of a C$_{128}$ polymer as functions of the system's temperature and pressure, at temperatures from 165 to 413 K and pressures from one atmosphere to 3.5 kbar.  Simulation methods duplicated those in earlier papers by Tsolou, et al.\cite{ROUSEtsolou2006a,ROUSEtsolou2006b}.  The authors first examined the time autocorrelation functions $\Phi_{ppii}(t)$ of the Rouse mode amplitudes for $p$ having various values in the range $(1,64)$.  The  $\Phi_{ppii}(t)$  were all found to decay as stretched exponentials in time, leading to a set of values for $\tau_{p}$ and $\beta_{p}$.  The stretched exponentials were also characterized via their total correlation times
\begin{equation}
    t_{p} = \frac{\Gamma(1/\beta_{p})}{\beta_{p}} \tau_{p}.
\label{eq:totaltime}
\end{equation}

Tsolou, et al.\cite{ROUSEtsolou2008a} found that the $t_{p}$ depend on temperature via a modified Vogel-Fulcher-Tamman equation
\begin{equation}
    t_{p}(T) = t_{p0} \exp\left(\frac{D_{p} T}{T-T_{o}}\right).
    \label{eq:tpTVFT}
\end{equation}
Here $T$ is the absolute temperature, $t_{p0}$ and $D_{p}$ are fitting parameters, and $T_{o}$ is a characteristic temperature. Over the range of temperatures that were examined in the simulations, the temperature dependences of the $t_{p}$ do not depend markedly on $p$.  Pressure dependences of the $t_{p}$ were obtained at temperatures 310 and 413 K.  The $t_{p}$ increase exponentially with increasing pressure $P$.  Viewed graphicly, the effect of $P$ on $t_{p}$ does not depend a great deal on $p$.  The authors define an activation volume for the $t_{p}$ via
\begin{equation}
    \Delta V_{p}(P) = R T \left(\frac{\partial \ln (t_{p})}{\partial P} \right)_{T}.
\label{eq:activationvolume}
\end{equation}
$\Delta V_{p}$ decreases roughly by two-fold between $p=1$ and $p=64$, and for smaller $p$ is modestly smaller (byless than 10\%) at the higher than at the lower temperature.

Tsolou, et al.\cite{ROUSEtsolou2008a} also calculated the single-chain intermediate structure factor\cite{ROUSEharmandaris2003a,ROUSEtsolou2005a}
\begin{equation}
    g^{(1)}(q,t) = \sum_{n, m=1}^{N}\sin(q R_{nm}(t)/qR_{nm}(t)
    \label{eq:1cohSqt}
\end{equation}
in which $n$ and $m$ label two of the $N$ segments of a single chain, $q$ being the magnitude of the scattering vector and $R_{nm}(t)$ being the distance between segments $n$ and $m$ at two times separated by $t$.  $g^{(1)}(q,t)$ was found to be described by a stretched exponential in $t$.  The stretching exponent $\beta$  was reported to change with pressure and to decrease with increasing $q$.  The total correlation times for $ g^{(1)}(q,t)$ were found to decrease with increasing $q$ and to increase exponentially with increasing pressure. The activation volumes $\Delta V(T)$, as calculated from the pressure dependences of these total correlation times, decrease with increasing $q$.

Tsolou, et al.\cite{ROUSEtsolou2008a} calculated the single-chain incoherent scattering factor $S_{\rm inc}(q,t)$, which differs from $S(q,t)$ in that in equation \ref{eq:1cohSqt} the restriction $n=m$ is forced.  $S_{\rm inc}(q,t)$ decays as a stretched exponential in time.   $\beta$ was found to increase from $\approx 0.5$ to $\approx 0.9$ as $q$ was increased over the observed range.  The total correlation time $t_{c}$ decreased strongly with increasing $q$.  Lines to guide the eye, drawn as  $\tau \sim q^{-2/\beta}$ for smaller $q$ and $\tau \sim q^{-2}$ at larger $q$, are harmonious with the $q$ dependences of $t_{c}$ at multiple temperatures and a full range of pressures.  The peculiar $-2/\beta$ exponent is an artifact of the stretched-exponential form $\exp(-(t/\tau)^{\beta})$ used to parameterize $S(q,t)$.  If the parameterization had instead been
\begin{equation}
     S(q,t) = S(q,0)  \exp(- \Gamma t^{\beta})
     \label{eq:sqtstrexp}
\end{equation}
then for smaller $q$ the result would have been $\Gamma \sim q^{2}$, while for larger $q$ the form $\Gamma \sim q^{2 \beta}$ would have appeared as an approximant for the correct series developed in the chapter on scattering.

The Rouse model predicts that the relaxation time of the $p^{\rm th}$ mode should depend on $p$ as $p^{-2}$. As the model also predicts that modes relax as simple exponentials in $t$, not the stretched exponentials actually found, there is no theoretical basis for identifying the total correlation time with the Rouse time. Indeed, $p^{2} t_{p}$ is not independent of $p$; it instead decreases by about 30\% as $p$ is increased from 1 to 20. Tsolou, et al.,\cite{ROUSEtsolou2008a} use the observation to estimate a longest relaxation time and hence a model-dependent zero-shear viscosity for the system.  The inferred viscosity follows a Vogel-Fulcher-Tamman form.

Tsolou, et al.,\cite{ROUSEtsolou2010a} simulated melts of ring polyethylenes containing 24-400 carbon atoms at nominal temperature and pressure of 450 K and 1 atmosphere.  Simulations were made with a united atom model treating methylene units as single atoms, a harmonic bond-stretching potential, a harmonic-in-bond-angle bending potential, a bond-torsional potential, and a 12-6 Lennard-Jones potential for atoms separated by more than three bonds, using the r-RESPA algorithm\cite{ROUSEmartyna1996a}  for molecular dynamics.  As two of a large number of properties (most not considered in this chapter), they calculated $\langle (X_{p}(0))^{2}\rangle$ and $\langle X_{p}(0)X_{p}(t)\rangle$.  For $N/p^{2} \geq 2$, the mean-square amplitude $\langle (X_{p}(0))^{2}\rangle$ increased linearly in $N/p^{2}$.  At smaller $N/p^{2}$, the increase in $\langle (X_{p}(0))^{2}\rangle$ with increasing $N/p^{2}$ was much more rapid than linear in $N/p^{2}$, contrary to the Rouse model prediction of linear behavior at all $N/p^{2}$.  The time correlation functions $\langle X_{p}(0)X_{p}(t)\rangle$ were found to depend on $t$ as stretched exponentials in time. Comparison was made between the Rouse prediction $\tau \sim N^{2}/p^{2}$ and the $\tau_{p}$ calculated from the time integral of  $\langle X_{p}(0)X_{p}(t)\rangle$.  The prediction was sustained for $N/p \geq 30$. At smaller $N/p$, $\tau_{p}$ decreases two-fold as $N/p$ is reduced from 30 toward 1.  Rouse model predictions for the time correlation function and its zero-time value are therefore confirmed only for larger values of $N/p$.

Bulacu and van der Giessen\cite{ROUSEbulacu2005a} simulated the effect of bending and torsional potential energies on a polymer melt. Their polymer was a bead-spring model, with a 6-12 Lennard-Jones potential truncated at its minimum, bonds between beads represented with a FENE potential, a cosine harmonic bending potential
\begin{equation}\label{eq:ROUSEcosharm}
   V(\theta) = \frac{1}{2} k_{\theta} (\cos(\theta) - \cos(\theta_{0}))^{2}
\end{equation}
with $\theta_{0} = 109.5^{o}$, and a coupled bending-torsion potential
\begin{equation}\label{eq:ROUSEbendtorsion}
    V = k_{\phi} \sin^{3}(\theta_{i-1}) \sin^{3}(\theta_{i}) \sum_{n=0}^{3} a_{n} \cos^{n}(\phi_{i}).
\end{equation}
The torsion potential refers four beads in a row along a polymer, the two internal angles $\theta_{i-1}$ and $\theta_{i}$ formed by the two overlapping sets of three beads in a row, and the dihedral angle $\phi$.  The $a_{n}$ were determined by quantum calculations for $n$-butane as $(a_{0}, a_{1}, a_{2}, a_{3}) = (3, -5.9, 2.06, 10.95)$.  Simulations used the velocity-Verlet algorithm; temperature was held steady with a heat bath's random force and friction factor. $k_{\theta}$ and $k_{\phi}$ are stiffness coefficients. Their systems included up to 1000 chains with $5 \leq N \leq 250$.

Static properties examined include the mean-square end-to-end distance and the radius of gyration; these were calculated both for the entire chain and also for all of its sub-chains. With increasing chain stiffness, $\langle R^{2} \rangle/\langle R_{g}^{2} \rangle > 6$ was found. Histogram distributions of bond lengths, bending angles, and dihedral angles were reported.  The bead-bead radial distribution functions, same-chain, different-chain, and all-chains, were reported as linear plots.  Linear plots, while totally orthodox, can lose details of $g(r)$.   Whitford and Phillies have previously shown that the range of $g(r)$ in a Lennard-Jones fluid is, at lower temperatures, much longer than is sometimes assumed,\cite{ROUSEwhitford2005a,ROUSEwhitford2005b}, as made apparent by plotting $\log(|g(r)-1|)$ against $r$.   The dependence of mean-square Rouse amplitudes $\langle (X_{p}(0))^{2}\rangle$ on $p$ and the two chain stiffness parameters was examined for an $N=35$  polymer.  While increasing $k_{\theta}$ has little effect on $\langle (X_{p}(0))^{2}\rangle$, increasing $k_{\phi}$ considerably reduces $\langle (X_{p}(0))^{2}\rangle$  for larger values of $p$, by close to five-fold for $p=20$.

Dynamic properties studied include the polymer self-diffusion coefficient, which for the $N=40$ chain depends on the stiffness coefficients as $k_{\theta}^{-0.48}$ and $k_{\phi}^{-0.52}$.  $D$ depends on $N$ via a smaller-$N$ and a larger-$N$ power law.  The authors determined the break between the two power laws by maximizing the sum of the regression coefficients in fits of the two laws to the data.  With increasing chain stiffness, $D$ is reduced. With increasing chain stiffness, for shorter chains $D$ depends more strongly on $N$, the transition from short-to long-chain behavior moves to larger $N$, and the $N$-dependence of $D$ in the long-chain regime increases, finally attaining $D \sim N^{-2.2}$.  Time correlation functions $\langle X_{p}(t)X_{p}(0)\rangle$ were found to follow stretched exponentials in time, with $\beta $ decreasing with increasing $p$ and with an increase in chain stiffness.

Moreno and Colmenero\cite{ROUSEmoreno2008a} report an extended simulation of an A-B blend of bead-spring polymers.  The polymers were all shorter than the known nominal entanglement length, which is approximately 32 monomer beads.  Beads were connected by a FENE potential and a monomer-monomer potential
\begin{equation}
   V_{\alpha \beta}(r) = 4 \epsilon ((\frac{\sigma_{\alpha \beta}}{r})^{12} - 7 c^{-12} + 6 c^{-14} (\frac{r}{\sigma_{\alpha \beta}})^{2},
   \label{eq:morenopotential2008a}
\end{equation}
with $\epsilon = 1$, $c=1.15$, and $\sigma_{\alpha \beta}$ = 1.6, 1.3, or 1, respectively, in units of $\sigma_{BB}$ for $(\alpha, \beta) = (A,A),(A, B)$, or $(B,B)$, respectively, with a packing fraction 0.53 and a cutoff $\sigma_{\alpha \beta}$.  Temperatures ranged from 1.5 to 0.33 in different simulations. All chains were the same length, with $N$ between 4 and 21;  dynamic asymmetry appeared because $\sigma_{\alpha \beta}$ differed between the two types of chain, the A chains being larger and more numerous (70\% of the total).

Moreno and Colmenero calculated the Rouse amplitude time correlation functions $\Phi_{pq}(t) = \langle X_{p}(0) X_{q}(t) \rangle$.  For $p \neq q$, $|\Phi_{pq}(0)| \leq 0.1$ was found, which was taken to indicate that cross-correlations between Rouse amplitudes are small.  One notes that while the individual off-diagonal $C_{p,q}(0)$ are small, for any $p$ or $q$ there may be a respectably large number of them, so the total effect of the cross-correlations might not be negligible.   The $\Phi_{pp}(t)$ were found to be described by stretched exponentials in $t$. The stretching exponent $\beta$ depends weakly on $p$, tending to decrease with increasing $p$.   Considering chains with $N=10$, for the larger, more numerous A chains and $T \geq 0.6$, $\beta \in (0.8, 0.9)$  For the smaller, less numerous $B$ chains, with decreasing temperature $\beta$ fell smoothly from 0.9 or so to 0.3 or so.  That is, the relaxations are not the pure exponentials predicted by the Rouse model, but are closer to being single exponentials  for the larger $A$ chains .

The two authors also considered how the measured $\tau_{p}$ scaled with $N/p$.  For the A chains, except for the shortest-wavelength modes, $\tau_{p}$ scales approximately as $(N/P)^{2}$ at all temperatures studied.  For $N/p \leq 2$, these being the shortest wavelength modes, $\tau_{p}$ tended to be slightly larger than a scaling line prediction.  For the smaller, less numerous $B$ chains, a normalized $\tau_{p}$ scaled as $(N/p)^{x}$, with $x \approx 2.2$ at high temperature, but increasing to $x \approx 3.5$ at the smallest temperature studied.  $x \approx 3.5$ matches with the reptation prediction $x =3$, except as the authors note they are considering polymers that are too short to be entangled, indeed, polymers with as few as 4 monomer beads.  For both chains, $N \langle (X_{p}(0))^{2}\rangle$ scales as $(N/P)^{2.2}$, which is not far from the Rouse value $x=2$. For chains of length $N=15$, the authors computed the mean-square displacement $\langle (\Delta r(t))^{2} \rangle$ of the center beads of the B chains as a function of time at various temperatures.  For times shorter than the Rouse time, there is at each temperature a region in which $\langle (\Delta r(t))^{2} \rangle \sim t^{y}$, with a $y$ that decreases markedly as the temperature is lowered.  However, as the authors note, the entanglement crossover is reached by making polymer chains long, while here a crossover with the same appearance is reached by increasing the dynamic asymmetry, the value of $\sigma_{AA}/\sigma_{BB}$ between two components, neither of which is entangled. The authors note suggestions that these anomalous diffusive behaviors arise because density fluctuations around each chain become slow, but slowness ``\emph{may be induced by entanglement, but data reported here for the fast component suggest that this is not a necessary ingredient.}"  They emphasize that, with sufficient dynamic asymmetry, entanglementlike dynamics are observed even for model bead-spring tetramers.

Brodeck, et al.,\cite{ROUSEbrodeck2009a} report simulations of a polyethylene oxide melt, and comparison with inelastic neutron scattering studies. Simulations used Materials Studio 4.1 and Discover-3 (version 2005.1) with the COMPASS force field.  In addition to potential energy terms reflecting bond stretching, bond bending, and bond torsion, their potential energy calculations include the coordinate cross-coupling terms known to be essential for calculating infrared and Raman frequencies.  The polyethylene oxide oxygen has a significant partial charge, leading to Coulombic interactions. A Lennard-Jones 6-9 potential was used for the general nonbonded interaction.  The simulation cell included five polymer chains, each composed of 43 monomer units in a $24.7\text{\AA }$cubic cell at temperatures 400, 375, and 350 K. Brodeck, et al., make a series of tests of implications of the Rouse model, finding that the available scattering data on this system all agree with predictions of the Rouse model, namely (i) for $q \leq 0.6\text{\AA}^{-1}$, a characteristic relaxation time scales as $q^{-4}$; (ii) the incoherent scattering function depends on time as $\exp(-\alpha t^{1/2})$, at least for times longer than 2 pS, and (iii) the Rouse rate $W_{\rm eff}$ from simulations agrees with experiment.

The Rouse mode amplitudes $X_{p}(t)$, their mean-square values, and their temporal autocorrelation functions were determined for $ 1 \leq p \leq 43$.  A test of orthogonality found that $\langle X_{p}(0) X_{q}(0) \rangle$ was $\ll 1$ except for some large mode numbers. For each $p$, $\langle X_{p}(0) X_{p}(t) \rangle$ was described well by a stretched exponential in time.  For small $p$, $\langle X_{p}(0) X_{p}(t) \rangle$ followed Rouse model behavior, so that  $\langle (X_{p}(0))^{2} \rangle$ and the integrated average relaxation time $\tau_{p}$ each scaled as $p^{-2}$.  For larger $p > 8$, $\tau_{p}$ decreased  relative to the expected $p^{-2}$ dependence, finally reaching perhaps 2/3 of expected value, while, over the same range of $p$, $\langle (X_{p}(0))^{2} \rangle$ fell to a quarter of its expected value.  $\langle X_{p}(0) X_{p}(t) \rangle$ was never a simple exponential.  For $p=1$, the stretching exponent $\beta$ was as large as 0.9.  With increasing $p$, $\beta$ falls, reaching a minimum of 0.7 or so for $N/p$ close to 1, and then increasing slightly as $N/p = 1$ is reached.  $\beta$ is weakly temperature-dependent, especially at large $p$, increasing by a few percent as the temperature is increased.

Lahmar, et al.\cite{ROUSElahmar2009a}  extend the earlier work of Tzoumanekas, et al.\cite{ROUSEtzoumanekas2009a} to consider polymer dynamics of their polymer model.  In their model, individual beads represent 20-carbon backbone segments.  Atomistic simulations were then used to determine the bead-bead nearest-neighbor-intrachain and interchain radial distribution functions and the three-bead angular distribution function. From these the potentials of average force and thence via an iterative process the mean forces between polymer beads were determined.  Bead motions were described using dissipative particle dynamics, in which there is a frictional force along each bead-bead line of centers that is proportional to the bead-bead velocity along that line, and a corresponding thermal force that keeps the system in thermal equilibrum.  Because the polymer beads are soft and can interpenetrate, a short range segmental repulsion force adequate to greatly reduce chain crossing was superposed. Systems with 6 to 40 beads, i.e., 120 to 800 carbon atoms in the backbone, were then examined.  The center-of-mass diffusion coefficient was found to depend on bead length as $N^{-1.4}$ for the shortest chains and $N^{-2.3}$ for the longest chains, these values not being in agreement with the Rouse and reptation model predictions $N^{-1}$ and $N^{-2}$, respectively, though the latter is in reasonable agreement with experiments on polymer melts.   The end-to-end vector reorientation time also depends on chain length, its short- and long-chain dependences being in approximate agreement with the Rouse and reptation model predictions.

Lahmar, et al.\cite{ROUSElahmar2009a} also examined the Rouse-amplitude time correlation functions $\langle \mathbf{X}_{p}(0) \cdot \mathbf{X}_{p}(t) \rangle$, finding that these decay as stretched exponentials in time, not the pure exponentials required by the Rouse model.  The stretching exponent $\beta$ depends on $p$ and on chain length.  For chains that cannot cross,  $\beta$ was smallest for $p$ in the range $N/2$ to $N/4$.   The authors propose that this length scale corresponds to the length scale of the network mesh as discussed by Tzoumanekas, et al.\cite{ROUSEtzoumanekas2009a}.  For the longest chain studied, $N=40$, $\beta$ decreased from 0.9 for the smallest and largest possible values of $p$ to 2/3 at its minimum. When chain crossing constraints were removed, $\beta$ was close to unity at small $p$ and decreased to 0.9 or so at large $p$. The mode relaxation times do not scale with $N$ and $p$ as predicted by the Rouse model.  In the absence of the segmental repulsion chain-crossing barrier, the mode relaxation times do not depend on chain length, but are slower than expected at large $p$.  In the presence of chain crossing constraints, for large $N$ the relaxation times only follow reptation model predictions for the first few values of $p$. Lahmar, et al., conclude that Rouse modes are not the system's normal modes, but do not claim that the system actually has normal modes of relaxation. Lahmar, etal., also considered rheological properties, to be discussed in the appropriate chapter.

Perez-Aparicio, et al.,\cite{ROUSEperezaparicio2010a} report a molecular dynamics simulation of poly(ethylene-\emph{alt}-propylene) based on a coarse-grained bead and spring model. Their coarse-grained potentials were the bead-bead potentials of average force determined by comparison with simulations made on shorter chains using an atomistic potential, with a single coarse-grained bead representing ten monomers, about half of the nominal entanglement length. Coarse-grained chains containing between 5 and 30 beads were then simulated.  A coarse-grained friction factor $f$ was set so that the long-time mean-square center-of-mass displacement of the coarse-grained short chains matched the long-time mean-square center-of-mass displacement of the atomistic chains. The short-time mean-square center-of-mass and bead displacements of the coarse-grained and atomistic chains do not match, apparently because for the coarse-grained chains a short-time frictional memory function has been replaced with a simple friction factor.

The authors then studied chain dynamics by calculating Rouse amplitudes and Rouse temporal autocorrelation functions $\Phi_{ppii}(t)$, which were found to follow stretched exponentials in time.  The dependence of the mean-square amplitude, $\langle (X_{p}(0))^{2} \rangle$, the stretching parameter $\beta_{p}$, an average relaxation time $\tau_{p}$ and a Rouse frequency $W_{\rm eff}$ on mode number $p$ and chain length $N$ were considered.  $W_{\rm eff}$ is the inverse of a nominal time needed for a bead to diffuse through the length of a statistical segment.   For $p \approx 1$, relaxations were very nearly exponential, with $\beta_{p} \approx 1$.  With increasing $p$, $\beta_{p}$ fell, declining to $\beta_{p} \approx 0.8$ for $N/p$  in the range 2-3.  At still larger $p$, $\beta_{p}$ increased again, reaching 0.95 or so for $N/p \approx 1$.  For $N/p > 5$,  $\langle (X_{p}(0))^{2} \rangle$ and $\tau_{p}$ followed the Rouse model predictions; at smaller $N/p$, $\langle (X_{p}(0))^{2} \rangle$ and $\tau_{p}$ are both substantially smaller than the Rouse predictions. The authors indicate their simulations are unrealistic at very small $N/p$, roughly $1 < N/p < 1.4$, this regime being much narriower than the $N/p <5$ regime in which Rouse behavior is not seen.  The Rouse frequency $W_{\rm eff}$ depends strongly on $N/p$, increasing perhaps fourfold as $p$ is increased from 1 to its upper limiting value $N-1$. These features are not consistent with the hypothesis that the Rouse model provides a valid description of polymer dynamics in the melt.

Perez-Aparicio, et al.,\cite{ROUSEperezaparicio2010a} repeat the warnings of Akkermans, Padding, and Briels\cite{ROUSEakkermans2000a,ROUSEakkermans2001a,ROUSEpadding2001c,ROUSEpadding2002a} that coarse-graining of atomic coordinates leads to friction forces, 'random' thermal forces, and can permit long polymer chains to pass through each other if appropriate precautions are not taken.  The rigorous statistico-mechanical representation of coarse-graining is provided by the Mori-Zwanzig formalism\cite{ROUSEphillies1999a}, in which the complete set of atomic coordinates is partitioned between variables retained for study and variables described as \emph{bath variables}, the latter being subject to a thermal averaging process that removes them from further consideration. The Mori-Zwanzig coarse-graining process introduces to the equations of motion a set of random, \emph{thermal} forces, a corresponding set of frictional forces that together with the thermal forces keep the system in thermal equilibrium, and a set of Mori memory kernels $\phi(t)$ that replace friction factors, namely
\begin{displaymath}
  - f \mathbf{v}(t)  \rightarrow  - \int_{-\infty}^{t} ds \, \phi(t-s) \mathbf{v}(s).
\end{displaymath}
Here $\mathbf{v}(t)$ is a bead velocity at time $t$.  One could in principle use simulations to recover the Mori memory kernels $\phi(t-s)$, as has been done in a different system by Phillies and Stott\cite{ROUSEphillies1995z,ROUSEphillies1995y}.

Perez-Aparicio, et al.,\cite{ROUSEperezaparicio2010a} explore possible paths for uniting their simulations with the Rouse model.  They note the possibility that the statistical segment length $b$ or the friction constant $f$ could depend on the mode, so that a mode dependence of $b$ or $f$  might explain some of their results.  In the context of the Rouse model, $b$ and $f$ are associated with particular bonds or beads, respectively, so it is unclear how a simple $p$-dependent form of these variables could be interpreted other than as a formal parameterization.  However, each $\Phi_{ppii}(t)$ samples $t$ with slightly different weightings for each time, and therefore represents a different averaging over a memory function $\phi(t-s)$, thus providing a mechanism that would lead to formal $p$ dependences of $b$ and $f$.  Some authors have noted that united-atom simulations show different short-time dynamics than do all-atom simulations.  Failure to treat friction with a memory function $\phi(t-s)$, as required by the Mori\cite{ROUSEmori1965a,ROUSEmori1973a} formalism, rather than with a simple friction factor $f$, would lead to this outcome. The extensive review by Jin, et al.,\cite{ROUSEjin2022a} discusses current results on applying approximations to the Mori-Zwanzig formalism to molecular dynamics in the presence of united atoms or coarse-graining.

Kalathi, et al.,\cite{ROUSEkalathi2014a} report simulations of melts of linear polymers, with individual chains containing between 10 and 500 beads. The objective was to analyze polymer motions in terms of Rouse modes, the hope being that the dynamics of longer- and shorter-wavelength (smaller and larger $p$) Rouse modes would reveal the motions of longer and shorter chain segments. As a justification for the study: Experimental and simulational evidence was cited for deviations from Rouse behavior at short distances and small times\cite{ROUSEperez2011a,ROUSEbrodeck2009a}. Temporal autocorrelation functions of Rouse mode amplitudes were found to decay as stretched exponentials in time, not the simple exponentials predicted by the Rouse model\cite{ROUSEshaffer1995b}, with mean-square amplitudes that fail to scale as the Rouse-predicted $p^{-2}$.  Also, the stretching parameter $\beta$ had previously been found by Padding and Briels to depend on $p$\cite{ROUSEpadding2001c,ROUSEpadding2002a}. Finally, Kalathi, et al., note the result of Likhtman\cite{ROUSElikhtman2012a}, that Rouse modes are sometimes substantially cross-correlated ($\langle X_{p}(t) X)_{q}(0) \rangle \neq 0$ for $p \neq q$).

In Kalathi, et al.,'s simulations, the bead-spring Kremer-Grest model was used, with a finitely extensible nonlinear-elastic potential for bonded beads, a Lennard-Jones interaction between all pairs of unbonded beads, and a bending potential between linked segments along each chain. This potential energy keeps chains from passing through each other. In some simulations, the potential energy was modified to permit chain crossing, the modified potential energy being such as not to change chain static properties significantly, thus testing for effects arising from topological interactions.  Rouse amplitude time correlation functions $\Phi_{ppii}(t)$ were evaluated for $p$ as large as 24 and four values for the chain stiffness. Simulations were executed out to sufficiently large $t$ that $\Phi_{ppii}(t)$ had decayed to essentially zero. Except at the earliest times, where in some systems for small $p$ the fitted curves pass above the data, their $\Phi_{ppii}(t)$ measurements were described well by stretched exponentials in time.

From the simulations, $\langle X_{p}^{2}(0) \rangle \sin^{2}(p \pi/2N)$ fails to be independent of $N/p$, contrary to the prediction of the Rouse model. This product does approach the characteristic ratio asymptotically for small $p$.  With decreasing $N/p$, the product $\langle X_{p}^{2}(0) \rangle \sin^{2}(p \pi/2N)$ decreases slowly until $N/p$ reaches 15 or so; at smaller $N/p$, the function decreases rapidly toward 0.5-0.8 at $N/p=2$.   The stretching parameter $\beta$ also depends on $N/p$, being roughly 0.9 at $p=1$, decreasing to a minimum $\beta \approx 0.5$ for $p \approx 10$, and then increasing to 0.75 or 0.8 for $N/p \leq 5$.  The minimum in $\beta$ is found for $N/p \approx N_{e}$, $N_{e}$ being an inferred entanglement length. $\beta$ also depends modestly on the bending constant in the force field.

Kalathi, et al., report results for a range of chain lengths, from $N=10$ to $N=500$.  At fixed $N/p$, $W^{\rm eff}$ is independent of $N$ for shorter chains, but for chains with $N \geq 150$ and $N/p > 15$ the authors report that $W^{\rm eff}$ decreases markedly with increasing $N/p$. $\beta$ also shows distinct behaviors for shorter and longer chains.  For $N \leq 100$, $\beta$ simply increases as $N/p$ is increased.  At larger $N$, $\beta$ shows a minimum at intermediate $N/p$.  Kalathi, et al., report the effective monomeric relaxation rate $W^{\rm eff}$ and the longest relaxation time.  $W^{\rm eff}$ increases by ten- to thirty-fold between small and large $p$.  Removing the chain crossing constraint considerably increases $W^{\rm eff}$, especially for longer chains, eliminates the dependence of $W^{\rm eff}$ on chain length, and eliminates the dip in $\beta$ at intermediate $N/p$. In the absence of the chain uncrossability constraint, the longest relaxation rate increased as $N^{2}$ for all chain lengths studied, rather than -- in the presence of chain uncrossability -- increasing more rapidly than $N^{2}$ at large $N$.

The central conclusion of Kalathi, et al.'s simulations is that the Rouse modes of their bead-spring polymers show one behavior if the chains are short and a different behavior if the chains are longer than some length $N_{e}$, but the short chain-long-chain difference disappears if the chains are enabled to pass through each other.  Their results provide direct evidence for the contribution of topological interactions (chain-crossing constraints) to the dynamics of long polymers.

Hsu and Kremer\cite{ROUSEhsu2016a,ROUSEhsu2017a} made molecular dynamics simulations of 1000 chains containing N =500, 1000, or 2000 beads at volume fraction 0.85, using a novel scheme for equilibrating a large assembly of long chains\cite{ROUSEzhang2014a}.  The chains were described by a bead-spring model\cite{ROUSEkremer1990b} with a FENE potential between bonded atoms, a Lennard-Jones potential between non-bonded atoms, and in different simulations a bend-bonding constant $k_{\theta}$ of 0 or 1.5. Molecular dynamics were executed using the ESPResSo++ package.   Averaging, especially of the stress tensor $\sigma_{\alpha \beta}$, was improved by making an ensemble as well as a time average, namely by averaging over ten independently-generated equilibrated systems as well as an extended time average.  The nominal distance $N_{e}$ between the hypothesized entanglement points was estimated from primitive path analysis and Green-Kubo determinations of viscoelastic properties to be 26-28 monomers, so the 1000-bead chain would nominally have contained three dozen entanglement lengths, and proportionately more or less for the 2000 and 500 bead chains.

Hsu and Kremer\cite{ROUSEhsu2016a} first calculated static correlation functions, including the distribution functions for the end-to-end distance and the radius of gyration, the mean-square distance between pairs of polymer beads as a function of the distance between them along the chain, the bond-bond orientation correlation function, and the static structure factor $S(q)$, thus identifying a good compromise value for the bond-bending constant.  They then develop several dynamic properties, beginning with the time dependence of mean-square displacements of inner monomers, of inner monomers with respect to the center of mass, and of the center of mass itself; comparison is made with power-law time dependences. They made a primitive path analysis based on cooling the chains while holding the ends fixed, and computed the stress relaxation modulus from the stress tensor.

Hsu and Kremer\cite{ROUSEhsu2017a} advanced to study the $\langle X_{p}(t)X_{p}(0)\rangle$ and $g^{(1)}(q,t)$.  For each $p$, $\Phi_{ppii}(t)$ was found to decay as a stretched exponential in $t$, relaxations being longer-lived at larger $N/p$.  The deviation from a simple-exponential decay was smaller for large $N/p$, i.e., for small $p$, and was less obvious at early times. For small $N/p$,  say $N/p \leq 25$, i.e., $p \geq N/25$,  $\Phi_{ppii}(t)$ was found to be independent of chain length. For larger $N/p$, $\Phi_{ppii}(t)$ was seen to depend on $N$, $\Phi_{ppii}(t)$ decaying less rapidly for the longer chains.   The parameters of the stretched exponential were found to depend on $N/p$.  The stretching exponent $\beta$  decreases from 0.8 or so at small $N/p$ to a minimum near 0.5 for $N/p$ near 25 or 80 (depending on the chain stiffness), and then increases to 0.6 or so for large $N/p$. The location of the minimum in $\beta$ is independent of the polymer length. The value of $N/p$ for which $\beta$ reaches its minimum is approximately the same as the value of $N/p$ above which $\Phi_{ppii}(t)$ begins to depend on $N$.  The average time constant $\tau$ increases through eight orders of magnitude as $N/p$ is increased from 1 to 1000. Hsu and Kremer compare $\tau$ to power laws $(N/p)^{x}$, using theoretical model values $x =2$ for small $N/p$ and $x =3.4$ for large $N/p$.  Regions where these predictions are approximately correct are indeed found, though, at large $N/p$, $\tau$ tends to roll off to below the $x=3.4$ power law curve.

Finally, Hsu and Kremer\cite{ROUSEhsu2017a} discuss the coherent and incoherent structure factors $g^{(1)}(q,t)$ and $g^{(1s)}(q,t)$.  $g^{(1s)}(q,t)$ is the self part of the sum that defines $g^{(1)}(q,t)$. Their analysis is based in the Gaussian approximation, e.g.,
\begin{equation}\label{eq:ROUSEsqtincoherent}
    g^{(1s)}(q,t) = \left\langle \sum_{i=1}^{N} \exp(\imath \mathbf{q} \cdot (\mathbf{r}_{i}(t) - \mathbf{r}_{i}(0)))\right\rangle \equiv  \sum_{i=1}^{N} \exp(-\frac{q^{2}}{6} \left\langle (\mathbf{r}_{i}(t) - \mathbf{r}_{i}(0))^{2}\right\rangle)
\end{equation}
where $\left\langle (\mathbf{r}_{i}(t) - \mathbf{r}_{i}(0))^{2}\right\rangle$ is the mean-square displacement of bead $i$ during time $t$.   Hsu and Kremer report $\log(g^{(1)}(q,t)$ and
$\log(g^{(1s)}(q,t))/q^{2}$  as functions of time for a half-dozen values of $q$.  If the Gaussian approximation were correct, the latter quantity would be independent of $q$.  Contrariwise, as discussed in the chapter on scattering, if $\log(g^{(1s)}(q,t))/q^{2}$ has a pronounced $q$-dependence, this quantity would be determined by the all moments $(\mathbf{r}_{i}(t) - \mathbf{r}_{i}(0))^{2n}$, $n \geq 1$ of the bead displacement, not just the mean-square bead displacement.  Indeed, Hsu and Kremer report that at short times  $\log(g^{(1s)}(q,t))/q^{2}$ is independent of $q$, so at short times the Gaussian approximation is correct and $\log(g^{(1s)}(q,t))/q^{2}$ gives the mean-square bead displacement.

At times longer than the hypothesized entanglement time $\tau_{e}$,  $\log(g^{(1s)})/q^{2}$  depends markedly on $q$, so in the hypothesized reptation regime the Gaussian approximation fails, in which case $\log(g^{(1s)}(q,t))/q^{2}$ does not reveal the mean-square displacement of the beads and does not provide a test of the existence of the proposed $t^{1/4}$ motional regime.   Hsu and Kremer's log-log plots of $g^{(1)}(q,t)$ against $q^{2}t^{1/2}$ show that $\log (g^{(1)}(q,t)/q^{2}$ is also non-trivially dependent on $q$, so in the hypothesized reptation regime the Gaussian approximation fails and  $g^{(1s)}(q,t)$ is no longer determined by the mean-square bead displacements. The hypothesized power laws from the reptation model were drawn as tangent lines to some of the $\log(g^{(1s)}(q,t))/q^{2}$ curves, transitions between the hypothesized motional regimes being estimated as points where the power-law lines intersect.  From the estimated times, an estimate of the entanglement length $N_{e}$ was obtained, this length being approximately in agreement with $N_{e}$ estimated from primitive path analysis or from melt viscoelastic properties.

Goto, et al.\cite{ROUSEgoto2021a} used the Kremer-Grest bead-spring model to study melts of linear and ring molecules.  Their calculations incorporated 10,000 beads, linked in different simulations into molecules having between 5 and 400 linked beads, beads interacting via FENE, bending angle, and Lennard-Jones potentials. The temporal autocorrelation functions of the Rouse amplitudes were calculated. For each molecular weight and $p$, $ \langle  X_{p}(0) X_{p}(t)\rangle$  was found to be described well by a stretched exponential in time. The stretching exponent $\beta$, the mean-square amplitude $\langle (X_{p}(0))^{2}\rangle$, and the relaxation time $\tau$ were all found to depend on the mode number $p$ and the length $N$ of the polymer chain as functions of the unified variable $N/p$.  For ring polymers, $\beta$ was never larger than 0.8, with a local minimum of 0.7 near $N/p \approx 4$, a return to 0.8 for $N/p \approx 20$, and a tendency to decrease at larger $N/p$, contrary to the Rouse model prediction $\beta =1$ at all $N/p$.  According to the Rouse model, the normalized Rouse amplitude $\sin^{2}(\pi p/N) \langle (X_{p}(0))^{2} \rangle$ should be a constant independent of $N/p$.  For both rings and linear chains, in Goto, et al.'s simulations this quantity instead increases roughly five-fold between $N/p \approx 1$ and $N/p \approx 30$; for ring polymers at $N/p > 30$, this quantity clearly declines again with increasing $N/p$. $\tau$ increases with increasing $N/p$, approximately as a power law in $N/p$,  the power being close to 2 at smaller $N/p$ and larger, especially for the linear polymers, at large $N/p$.  The exponent of the power law visibly increases smoothly with increasing $N/p$, without a sharp transition in its value.  In summary, the Rouse model predictions for the form of $\langle (X_{p}(0))^{2}\rangle$ are not consistent with these simulations.

Goto, et al.\cite{ROUSEgoto2021a} also examined $P(r,t)$,  the probability distributions for the displacement of single beads and for chain centers of mass.  For ring polymers, $P(r,t)$ remains nearly Gaussian at all times.  For linear polymers, at larger times $P(r,t)$ deviates from the expected Gaussian form, becoming broader as $t$ increases. Characterizing the deviation from a simple Gaussian with a non-Gaussian parameter $\alpha_{2}(t)$
\begin{equation}
    \alpha_{2}(t) = \frac{3 \langle (R(t))^{4} \rangle - 5 \langle (R(t))^{2} \rangle^{2}  }{5 \langle (R(t))^{2} \rangle^{2}}
    \label{eq:RousennonGaussian}
\end{equation}
$\alpha_{2}(t)$ for monomers in linear chains has a small short-time peak, roughly the same size for all chain lengths, and a long-time peak whose size increases perhaps linearly with the chain length. For monomers in ring polymers, only the small short-term peak is apparent.

Roh, et al.,\cite{ROUSErohej2019a} simulated ring and linear polyethylenes that had short-chain branches spaced along their lengths. The polyethylene backbones each contained 400 carbon atoms, with 33 side chains, each five carbon atoms in length; as a control simulations were made of polymers having the same backbones but no side chains. Simulated systems contained $54-70$ molecules held at constant temperature and pressure. Molecules interacted via TraPPE united-atom potentials\cite{ROUSEtuckerman1992a}, motions being calculated with the r-RESPA algorithm\cite{ROUSEmartin1998a,ROUSEmartin1999a}.  The addition of short chains to the polymers renders the molecules more compact.

Rouse mode amplitudes $X_{p}(t)$ were computed.  For linear chains, the equilibrium average $\langle (X_{p}(0))^{2} \rangle$ were linear in $N/p^{2}$ for large $N/p^{2}$, but, for $N/p^{2} \leq 10$,  $\langle (X_{p}(0))^{2} \rangle$ fell below the linear line.  For ring polymers,  $\langle (X_{p}(t))^{2} \rangle$ had a linear dependence on $N/p^{2}$, with a slight drop-off from that dependence in the limits of small or large $N/p^{2}$.  Roh, et al., also examined the temporal autocorrelation functions  $\langle X_{p}(0 )X_{p}(t) \rangle$.  In all cases examined,  $\langle X_{p}(0 )X_{p}(t) \rangle$ deviates markedly from a simple exponential decay, having instead the form of a stretched exponential in time.  Writing the stretched exponential as $\exp(-(t/\tau_{p})^{\beta})$, $\tau_{p}$ was found to scale as $(N/p)^{2}$.

A few studies have used very different bead-bead potential energies.  Smith, et al.\cite{ROUSEsmith1995a,ROUSEsmith1996a} used a tangent hard-sphere polymer model, in which the bond length between adjoining spheres was allowed to vary over a narrow range. Their polymer chains had any of six lengths between 8 and 192 spheres, at volume fractions 0.3, 0.4, or 0.45. Systems contained 32 or 64 chains in a cell with periodic boundary conditions. The simulations covered five orders of magnitude in time. Smith, et al., studied the behavior of static and dynamic properties, including chain dimensions, Rouse mode amplitudes, chain mean-square displacements, fluctuations in the chain end-to-end vector's length and orientation, evidence for knot formation, dynamic structure factors, and Rouse-mode time autocorrelation functions.  Smith, et al., note that reptation and the tube model predict that, at times shorter than the entanglement time and chain segments shorter that the mean number of beads between two entanglements, the Rouse model should be applicable, but at long times the Rouse mode relaxation rates should increase in proportion to the number of times that a typical chain is entangled.

From the amplitudes $X_{p}(t)$ of the Rouse mode, Smith, et al.,\cite{ROUSEsmith1995a,ROUSEsmith1996a} calculated the normalized temporal autocorrelation functions $\Phi_{ppii}(t) = \langle X_{p}(0) X_{p}(t) \rangle/\langle (X_{p}(0))^{2} \rangle$.  These were plotted against the reduced times $t p^{2}/N^{2}$, where $p$ is the mode number and $N$ is the chain length.  For a 32-bead polymer, at all volume fractions $\Phi_{ppii}(t)$ is very nearly exponential.  Especially at larger polymer concentration, the higher-order modes ($2 \leq p \leq 6$), deviate from simple-exponential behavior, their relaxations being slower-than exponential at longer times,  with relaxation rates scaling linearly in $p^{2}$.  For a 192-bead polymer, relaxations are markedly non-exponential, the deviation from single-exponential behavior being most notable at the largest polymer volume fraction.  In contrast to the 32-bead polymer, for the 196-bead polymer the non-exponential behavior is most prominent for small $p$. Also, for the 196-bead polymers the mode relaxation rates are not linear in $p^{2}$.  At $\phi = 0.45$, $\Phi_{ppii}(t)$ gains a pronounced shoulder, and then decays very rapidly toward zero. The Rouse model prediction of exponential decay of $\Phi_{ppii}(t)(t)$  is thus not sustained for longer polymer chains.

Smith, et al., propose to have seen Rouse behavior, based on the observations $D \sim N^{-1}$ for shorter polymers and mean-square displacement proportional to $t^{1/2}$ at shorter times.  At intermediate times they find mean-square displacements $\sim t^{0.28}$ or $t^{0.31}$.  $t^{1/4}$ behavior was not observed.  Their $D$ against $N$ plot shows for each polymer volume fraction a smooth curve approaching being tangent to $D \sim N^{-1}$ at small $N$ and to $D \sim N^{-2}$ at large $N$. They also note an anomaly, namely, following the second $t^{1/2}$ regime in the  mean-square displacements, they observe a small plateau in the time dependences of the mean-square displacement and the end-to-end vector time correlation function.  They propose that the plateau corresponds to a time-scale separation between two types of entanglements, namely tight local knots and surrounding chains functioning as fixed obstacles.

Can Rouse coordinates and modes emerge naturally from an analysis of Brownian or molecular dynamics simulations?  This question was approached by Wong and Choi\cite{ROUSEwong2019a}, who consider a non-linear bead-spring model in which individual terms of the linked-bead potential have a form
\begin{equation}\label{eq:ROUSEnonzerolength}
   U_{ij} = \frac{1}{2}  k (|\mathbf{r}_{i} -\mathbf{r}_{j}| -b)^{2}.
\end{equation}
Here $k$ is a force constant, $\mathbf{r}_{i}$ is the position of bead $i$, and $b$ is a constant.  This potential energy term has a minimum when beads $i$ and $j$ are separated by the distance $b$. In the original Rouse model, $b=0$; when in their rest positions, all beads of a Rouse chain are at the same location. For their Brownian dynamics simulations,  Wong and Choi integrated their equations of motion with the Newton-Gauss method.  Wong and Choi also made molecular dynamics simulations of chains in a melt, using the TraPPE\cite{ROUSEmartin1998a,ROUSEmartin1999a} force field with GROMACS\cite{ROUSEabraham2015a} as an integrator. For the molecular dynamics simulations, individual chains had 30-73 beads, the number of chains being fixed at 40 for simulations of longer chains.

For each molecule and time step, Wong and Choi then calculated, separately along each Cartesian coordinate, the distance $q'_{i}$ for each bead $i$ from the chain center of mass. They then formed and averaged the $N \times N$ single-time correlation matrices $C_{i,j}$
\begin{equation}\label{eq:ROUSEcorrelationmatrix}
    C_{i,j} = \frac{1}{S}  \sum_{t=0}^{S} q'_{i}(t)q'_{j}(t),
\end{equation}
$t$ here being the label on the $S$ time steps. Proper Orthogonal Decomposition was then used to obtain dominant eigenvectors of $C_{i,j}$.  Simulations of single chains using Brownian dynamics and simulations of chain melts using molecular dynamics yielded similar forms for the eigenvectors of $C_{i,j}$.  Wong and Choi examined linear, ring, and star polymers.   The dominant eigenvectors of the linear polymer closely resemble the eigenvectors of the Rouse model.  The temporal autocorrelation functions of the mode amplitudes were then evaluated, finding that the autocorrelation functions decay as stretched exponentials in time, with values of $\beta$ close to unity. For ring and star polymers, two or three, respectively, of the relaxations were found to be nearly degenerate in their relaxation times.  The temporal cross-correlation functions were not evaluated. The integrated average relaxation time for the modes scales differently for the Brownian dynamics and the molecular dynamics simulations, namely $\tau \sim (N/p)^{2}$ for the Brownian dynamics simulations but $\tau \sim (N/p)^{2.3}$ for the molecular dynamics simulations.

\section{Conclusions\label{ROUSEconclusions}}

Summarizing from many papers, with respect to polymer dynamics and the Rouse model.

First, the Rouse coordinates are just that, coordinates, generated from the Cartesian coordinates of the model polymer beads by means of a discrete Fourier transform.  This is a purely mathematical replacement of one representation of the bead positions with another, and has no physical content.  Correspondingly, a challenge to the validity of the Rouse coordinates \emph{as coordinates} is a challenge to the validity of Fourier's theorem, so the challenge is highly unlikely to be correct.

However, the Rouse \emph{coordinates} are sometimes also interpreted as the Rouse \emph{modes}, which within the Rouse model of a single isolated polymer chain are predicted to have a series of physical properties. In a polymer melt, the Rouse modes do not have the predicted physical properties.  In particular:

1) The relaxation $\langle X_{p}(0)  X_{p}(t) \rangle$ of the temporal autocorrelation function of a single Rouse amplitude is a stretched exponential in time, not the pure exponential predicted by the Rouse model. At larger $p$ (shorter wavelength), the degree of stretching is more pronounced.

2) The mean-square amplitude of the Rouse modes $\langle X_{p}(0)  X_{p}(0) \rangle$ deviates from the model prediction, at least for $p > 3$.

3) The relaxation time $\tau_{p}$ of   $\langle X_{p}(0)  X_{p}(t) \rangle$ depends on $p$, but not as predicted by the Rouse model.

4) In one paper\cite{ROUSElikhtman2012a} where a cross-correlation function  $\langle X_{p}(0)  X_{q}(t) \rangle$, $p \neq q$, was evaluated, the cross correlation function at intermediate times became large.  Other authors have not duplicated this result.

5) Contrary to the Rouse model, under shear, the response of a chain is to rotate, not to distort affinely without rotation.

6) According to the Rouse model, bead displacements are driven by independent Gaussian random processes.  As a result, $g^{(1s)}(q,t)$ is accurately described by the Gaussian approximation.  Doob's theorem\cite{ROUSEdoob1942a} then guarantees that $g^{(1s)}(q,t)$ decays as a single exponential in time.  These predictions  are incorrect for polymer coils in the melt.

Now, it has been proposed that the deviation of $\tau_{p}$ from the Rouse predictions can be reduced by claiming that the bead friction factor $f$ is $p$-dependent.  The proposal is literally true, in the sense that $\tau_{p} \sim f^{-1}$ is the Rouse prediction, so by assigning $f$ a different value for each $p$ the model can predict any values whatsoever for the $\tau_{p}$.  However, the proposal makes no physical sense, namely a value of $f$ is associated with each bead, not with each mode.  When a bead moves it has no way to know which mode has been excited.  Furthermore, if two modes are excited simultaneously, each bead would be required to move with two different friction factors at the same time, which does not appear to make sense.

The above direct tests of the Rouse model correspond to the Rouse model's major predictions.  Without exception, the predictions are conclusively rejected by simulations.

There can be no doubt that the Rouse model is invalid in polymeric fluids.

\end{document}